\documentclass[aps,prl,reprint,groupedaddress,amsmath,amssymb,superscriptaddress]{revtex4-2}
\usepackage{graphicx}
\usepackage{float}
\usepackage{multirow}
\usepackage{natbib,twoopt}
\usepackage[breaklinks=true]{hyperref}
\usepackage{xcolor}
\usepackage{amssymb}
\usepackage{stackengine,graphicx}
\usepackage{amsmath}
\usepackage{hyperref}
\usepackage{color}
\usepackage{stackengine}
\usepackage{subfigure}
\usepackage{verbatim}
\usepackage{mathtools}
\usepackage{braket}
\usepackage{bm}


\newcommand{\RNum}[1]{\uppercase\expandafter{\romannumeral #1\relax}}

\newcommand{\balancecolsandclearpage}{%
	\close@column@grid
	\cleardoublepage
	\twocolumngrid
}

\begin{document}

\title{Axion electrodynamics without Witten effect in metamaterials}

\author{Eduardo Barredo-Alamilla}
\email{eduardo.barredo@metalab.ifmo.ru}
\affiliation{School of Physics and Engineering, ITMO University, St. Petersburg, Russia}
\author{Daniel A. Bobylev}
\email{daniil.bobylev@metalab.ifmo.ru}
\affiliation{School of Physics and Engineering, ITMO University, St. Petersburg, Russia}
\author{Maxim A. Gorlach}
\email{m.gorlach@metalab.ifmo.ru}
\affiliation{School of Physics and Engineering, ITMO University, St. Petersburg, Russia}

\begin{abstract}
Artificial media provide unique playground to test fundamental theories allowing one to probe the laws of electromagnetism in the presence of hypothetical axions. While some materials are known to realize this physics, here we propose the nonlocal extension of axion electrodynamics. Compared to the usual axion case, the suggested metamaterial features similar optical properties including Kerr and Faraday rotation. However, the external sources in this structure do not induce dyon charges eliminating well-celebrated Witten effect. We put forward the design of such nonreciprocal non-local metamaterial and discuss its potential applications.
\end{abstract}

\date{\today}

\maketitle


\textit{Introduction.}~--- The axion field was proposed to resolve strong CP problem in quantum chromodynamics (QCD)~\cite{PecceiQuinn1977,Weinberg1977,Wilczek1978}. Since then, its fundamental quanta, axions, attract the interest of the physics community including such areas as high-energy physics, cosmology and string theory~\cite{MARSH2016,Svrcek_2006}. Since axion is anticipated to be extremely light and weakly interacting with the ordinary matter, this hypothetical particle or its generalizations known as axion-like particles (ALPs) are viewed as promising candidates for cosmological dark matter~\cite{Millar2023,Adair2022,marsh2017}. 


The coupling between axions and electromagnetic fields is captured by the axion electrodynamics~\cite{Sikivie1983,Wilczek1987} which features an additional term in the Lagrangian density $\mathcal{L}_\theta \sim \kappa a(x)\mathbf{E}\cdot\mathbf{B}$, where $a(x)$ is the axion field and $\kappa$ is the axion-photon coupling. An important prediction of the theory is the Witten effect~\cite{Witten1979}: magnetic charge in a vacuum bubble surrounded by the axion medium gives rise to the effective dyon charges, i.e. the combination of electric and magnetic monopoles fields [Fig.~\ref{axionvspsicase}(a)].

Axion electrodynamics has found a number of realizations in condensed matter physics~\cite{Nenno2020,Sekine2021} and, recently, in metamaterials~\cite{shaposhnikov2023,Prudencio2023,Jazi2023}, since low-energy effective description of those structures yields the axion Lagrangian. Alternatively, the same physics can be viewed through the prism of the constitutive relations. In such case, the equations of axion electrodynamics correspond to the special non-reciprocal bianisotropic medium with the constitutive relations of the form $\mathbf{D}= \varepsilon\,\mathbf{E}+\theta(x) {\mathbf B}$ and $\mathbf{H}= \mu^{-1}\,\mathbf{B}-\theta(x) {\mathbf E}$, known in the photonics community as Tellegen media. Evidently, $\theta$-term in these equations preserves $\mathcal{PT}$-symmetry, while breaking $\mathcal{P}$ and $\mathcal{T}$ symmetries separately.

In this Letter, we make the next conceptual step showing that the metamaterial platform allows one not only to implement synthetic axion fields, but actually go beyond that probing nonlocal extensions of axion electrodynamics and associated exotic physics. While the suggested metamaterial responds to plane wave excitation similarly to the standard axion case, its interaction with the external localized sources is fundamentally different. In particular, Witten effect in this structure further termed $\psi$-metamaterial vanishes. Below, we explore the physics of $\psi$-medium, put forward its possible experimental implementation and discuss potential applications.


\textit{$\psi$-electrodynamics.}~--- We investigate a metamaterial described by the constitutive relations:
\begin{figure}[h!]
	\centering
	\includegraphics[width=0.48\textwidth]{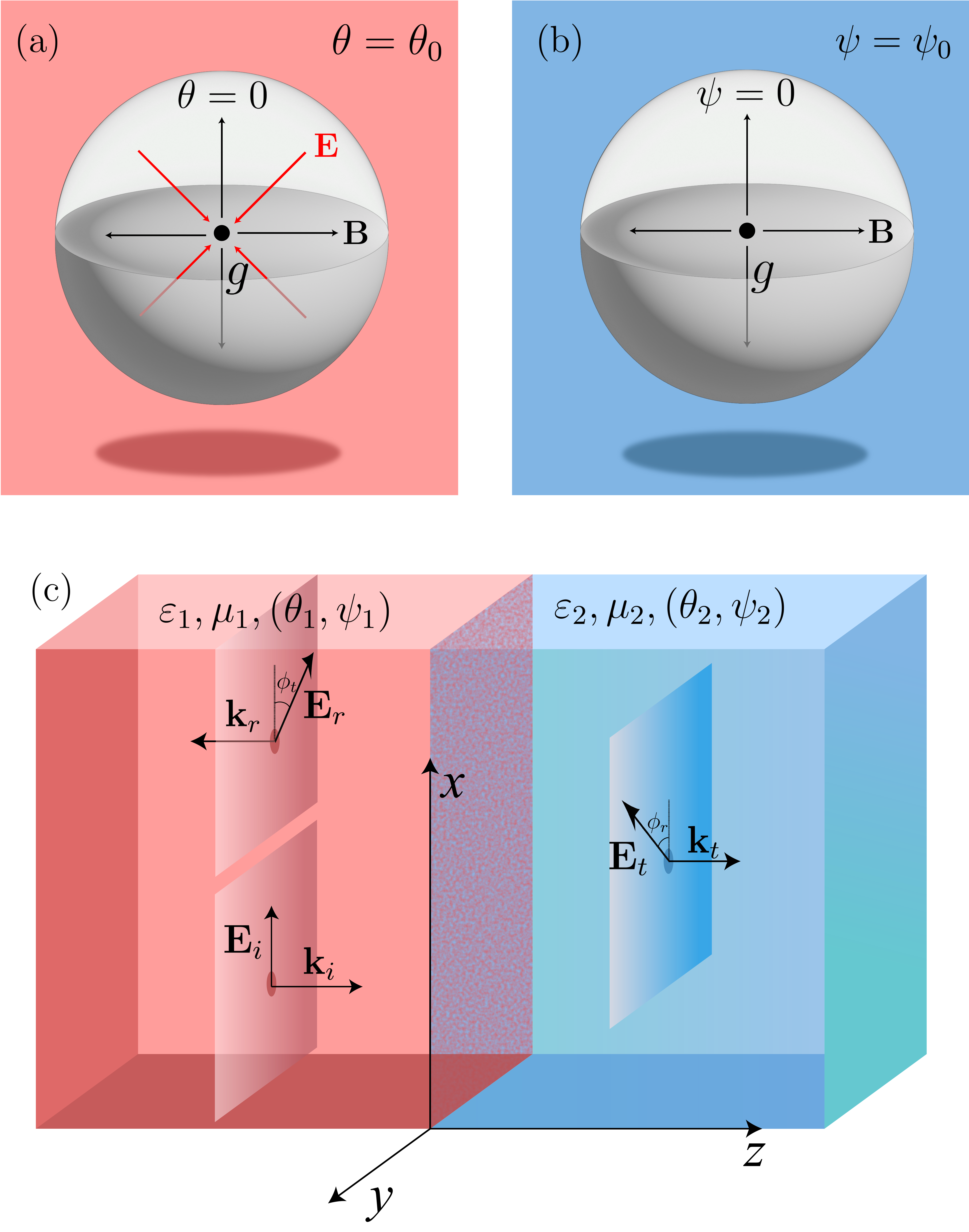}
	\caption{Comparison of the electromagnetic properties between the $\theta$- and $\psi$-response. (a) Witten effect: Electric field induced by the magnetic monopole $g$ in a vacuum bubble surrounded by the constant axion field $\theta_0$. Black and red arrows denote force lines of the magnetic and electric field, respectively. (b) In the case of $\psi$-electrodynamics there is no induced electric field, i.e. no Witten effect. (c) Light transmitted and reflected at $\theta$- or $\psi$-interface experiences the rotation of the polarization plane.}
	\label{axionvspsicase}
\end{figure}
\begin{eqnarray}
\mathbf{D}_{\psi} & = &\varepsilon \mathbf{E} + \psi\nabla^2 \mathbf{B}, \label{auxiliarD} \\
\mathbf{H}_{\psi} & = &\mu^{-1} \mathbf{B} - \psi\nabla^2 \mathbf{E}. \label{auxiliarH}
\end{eqnarray}
The $\psi$-term in these equations shares with the axion field $\theta$ the same symmetry properties with respect to $\mathcal{P}$ and $\mathcal{T}$; explicit breaking of $\mathcal{T}$ symmetry indicates nonreciprocity of the medium. However, differently from the axion case, the constitutive relations contain spatial derivatives of the field pointing towards electromagnetic nonlocality. If formulated in terms of the effective Lagrangian, this corresponds to the introduction of higher-order derivatives~\cite{Supplement}. Such feature of the Lagrangian is not common in high energy physics. However, as we prove below, it is readily attainable in metamaterials context.


Using Eqs.~\eqref{auxiliarD}, \eqref{auxiliarH} with spatially varying $\psi$ and assuming CGS system of units, we present Maxwell's equations in the form
\begin{eqnarray} 
\nabla \cdot (\varepsilon\mathbf{E}) & = & 4\pi \rho-\nabla \psi \cdot \nabla^2\mathbf{B} \label{psieqs1},\\
\nabla \times \left( {\mu}^{-1}\mathbf{B}\right) -\frac{1}{c}\frac{\partial (\varepsilon\mathbf{E}) }{\partial t} & = & \frac{4\pi}{c} \mathbf{J} + \nabla \psi \times \nabla^2\mathbf{E} \label{psieqs2}, \\
\nabla \cdot \mathbf{B} & = & 0 \label{psieqs3}, \\
\nabla \times \mathbf{E}+\frac{1}{c}\frac{\partial \mathbf{B}}{\partial t} & = & 0,
\label{psieqs4}\end{eqnarray}
where $\varepsilon$ is the permittivity and $\mu$ the permeability of the metamaterial. Extra terms arising in Eqs.~(\ref{psieqs1}),(\ref{psieqs2}) can be regarded as effective charge and current densities: $\rho_\psi = -\frac{1}{4\pi} \nabla \psi \cdot \nabla^2\mathbf{B}$, $\mathbf{J}_\psi = \frac{c}{4\pi} \nabla \psi \times \nabla^2\mathbf{E}$. Thus, if $\psi$ is constant, effective charge and current densities vanish yielding the conventional Maxwell's equations.

\begin{figure}[ht!]
	\centering
	\includegraphics[width=0.4\textwidth]{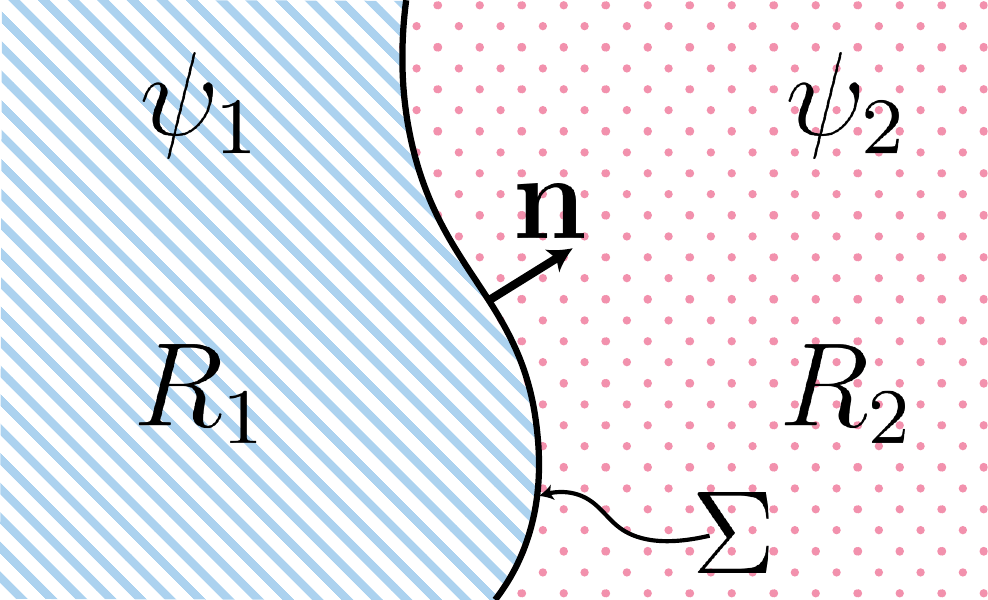}
	\caption{
	    Interface between two distinct $\psi$-metamaterials with different permittivity and permeability.
	}
	\label{Fig1}
\end{figure}

As a simplest nontrivial example, we consider a piecewise-constant function $\psi$ that takes the constant value $\psi_1$ in a region $R_1$ and $\psi_2$ in a region $R_2$.
%
%
The gradient of the piecewise function $\psi$ is
\begin{equation}
    \nabla \psi = -\tilde{\psi}\delta(\Sigma)\hat{\mathbf{n}}, \qquad \Tilde{\psi} = \psi_1 - \psi_2,
\label{psigradient}\end{equation}
where $\Sigma$ is the boundary of the regions $R_1$ and $R_2$ and $\hat{\mathbf{n}}$ is a unit vector normal to the boundary pointing from $R_1$ to $R_2$ [Fig.~\ref{Fig1}]. Hence, similarly to the axion case~\cite{MartinRuiz2015}, the only modification of Maxwell's equations occurs at the boundary due to the additional boundary currents. In the absence of externally introduced charges and currents the boundary conditions read~\cite{Supplement} 
\begin{align}
   [\mathbf{D}]_n & = \tilde{\psi}\, \nabla^2\mathbf{B}\cdot \hat{\mathbf{n}}|_\Sigma, & [\mathbf{H}]_\parallel & = -\tilde{\psi}\, \nabla^2\mathbf{E}\times \hat{\mathbf{n}}|_\Sigma, \label{PsiBC1} \\
   [\mathbf{B}]_n & = 0, & [\mathbf{E}]_\parallel & = 0, \label{PsiBC2}
\end{align}
%
where $\mathbf{D}=\varepsilon\mathbf{E}$ and $\mathbf{H}= \mu^{-1} \mathbf{B}$,  $[\mathbf{A}]_n = (\mathbf{A}^{(2)}-\mathbf{A}^{(1)})\cdot \hat{\mathbf{n}}$ and $[\mathbf{A}]_\parallel = (\mathbf{A}^{(2)}-\mathbf{A}^{(1)})\times\hat{\mathbf{n}}$. Here, $\mathbf{A}^{(1)}$ and $\mathbf{A}^{(2)}$ are the fields in the regions $1$ and $2$ close to the boundary $\Sigma$. To make the problem well-defined, we take the half-sum of the Laplacians from both sides of the interface in Eq.~\eqref{PsiBC1}.


\textit{Electromagnetic waves in $\psi$-metamaterial.}~--- Having the set of boundary conditions, we examine the propagation of electromagnetic waves through the planar interface $z=0$ between two metamaterials with $\psi_1$, $\varepsilon_1$, $\mu_1$ for $z<0$ and $\psi_2$, $\varepsilon_2$, $\mu_2$  for $z>0$, respectively, using the conventional ansatz for the plane wave solutions:
\begin{equation}
    \begin{pmatrix} \mathbf{E}(t,\mathbf{r}) \\ \mathbf{B}(t,\mathbf{r})
    \end{pmatrix} = \begin{pmatrix} \mathbf{E}(\omega,\mathbf{k}) \\ \mathbf{B}(\omega,\mathbf{k})
    \end{pmatrix} e^{-i\omega t + i \mathbf{k}\cdot\mathbf{r}}.
\end{equation}
The fields of the plane wave $E_{\alpha,\beta}(\omega,{\bf k})$, $B_{\alpha,\beta}(\omega,{\bf k})$ are labelled by the two indices $\alpha$ and $\beta$. The first index $\alpha=i,r,t$ distinguishes incident, reflected and transmitted waves, while the second index $\beta=\bot,||$ denotes TE or TM polarization of the wave, respectively.

In the plane wave scenario, the Laplacian term in Eq.~\eqref{PsiBC1} acts as
\begin{equation}
\nabla^2 \xrightarrow{} -k_0^2\,\overline{\varepsilon\,\mu},
\end{equation}
where $k_0=\omega/c$ and $\overline{\varepsilon\,\mu} = \left(\varepsilon_1\mu_1 + \varepsilon_2\mu_2\right)/2$ is the average between the two media. As a result, the boundary conditions coincide with those in axion electrodynamics with the identification
\begin{equation}\label{Identification}
    \theta_1-\theta_2=-(\psi_1-\psi_2)\,k_0^2\,\overline{\varepsilon\,\mu}\:.
\end{equation}

As a result, the response of the metamaterial to the plane wave excitation is almost indistinguishable from the axion case. In particular, transmitted and reflected light will experience the rotation of the polarization plane, $\tan \phi_\alpha = E_{\alpha,\perp}/E_{\alpha,\parallel}$, known as Faraday and Kerr rotations, respectively:
\begin{equation}
    \tan \phi_t = \frac{ \tilde{\psi} \,k_0^2\,\overline{\varepsilon\,\mu}  }{\sqrt{\frac{\varepsilon_1}{\mu_1}}+\sqrt{\frac{\varepsilon_2}{\mu_2}}}, 
\end{equation}

\begin{equation}
    \tan \phi_r = \frac{2\sqrt{\frac{\varepsilon_1}{\mu_1}}\,\Tilde{\psi}\,k_0^2\,\overline{\varepsilon\,\mu} }{\frac{\varepsilon_2}{\mu_2}- \frac{\varepsilon_1}{\mu_1}+ \Tilde{\psi}^2\,k_0^4 \,\overline{\varepsilon \,\mu}^2 },
\end{equation}
where normal incidence is considered. These expressions coincide with those in the axion electrodynamics~~\cite{Qi2008,Chang2009,Tse2010,Maciejko2010} with the geometry of the problem depicted in Fig.~\ref{axionvspsicase}(c).

The parallel with the axion electrodynamics can be extended further by examining the surface waves at the interface of the two media with distinct values of $\psi$. Surface waves are superpositions of TE and TM polarizations with the dispersion given by 
\begin{equation}
    \left(\frac{q}{\mu_1}+\frac{p}{\mu_2}\right)\left(\frac{q}{\varepsilon_1}+\frac{p}{\varepsilon_2} \right) + \Tilde{\psi}^2\,k_0^4 \,\left(\overline{\varepsilon \,\mu}\right)^2 \frac{pq}{\varepsilon_1\varepsilon_2}=0,
\end{equation}
where $q^2  = \beta^2 -k_0^2 \mu_1\varepsilon_1 > 0$, $ p^2 = \beta^2 -k_0^2 \mu_2\varepsilon_2 > 0$, and $\beta$ is the propagation constant along the interface. This dispersion relation is also consistent with the axion electrodynamics~\cite{Maimistov2016}.

\textit{Witten effect.}~--- However, the situation changes when the metamaterial interacts with the localized sources. To illustrate this physics, we consider a magnetic monopole $g$ inside a spherical volume with the radius $a$ surrounded  by a $\psi$-metamaterial with nonzero and constant $\psi =\psi_0$. 

Because of the $\psi$-term in Eq.~(\ref{psieqs1}), the shell carries a charge density $-\nabla\psi \cdot \nabla^2\mathbf{B}/(4\pi)$. Since the field of the monopole is given by $\mathbf{B}= g\hat{\mathbf{r}}/r^2$, where $\hat{\mathbf{r}}$ is a unit vector in the radial direction, the modified Gauss law in Eq.~(\ref{psieqs1}) yields 


\begin{equation}
    \nabla\cdot\mathbf{E} = -4\pi g \psi_0 \delta(\mathbf{r}-\mathbf{a}) \frac{d}{dr}\delta(\mathbf{r}),
\label{monopolechargedensity}\end{equation}
where $\nabla \psi = \psi_0\delta(\mathbf{r}-\mathbf{a}) \hat{\mathbf{r}}$ from Eq. (\ref{psigradient}). The integration of Eq.~(\ref{monopolechargedensity}) yields a zero induced electric charge. Hence, in stark contrast to the conventional Witten effect, there are no induced effective dyon charges in the designed metamaterial [Fig.~\ref{axionvspsicase}(b)]. 

The absence of the Witten effect in $\psi$-metamaterial has further consequences. In particular, point magnetic dipole surrounded by a spherical shell made of $\psi$-metamaterial does not give rise to the combination of effective magnetic and electric dipoles~\cite{Supplement} contrary to the axion electrodynamics case~\cite{seidov2023}. Furthermore, the application of a constant electric field to a $\psi$-sphere does not induce magnetic dipole moments inside it as one could expect from the conventional axion case.

\textit{Engineering $\psi$-response.}~--- While the constitutive relations Eqs.~\eqref{auxiliarD},\eqref{auxiliarH} and the properties of the suggested material look exotic, we demonstrate below that the metamaterial platform can enable this sort of physics.

To come up with a suitable design, we recall that the Tellegen medium, i.e. artificial structure with the conventional axion response, was originally suggested as a composite where electric and magnetic dipoles are attached to each other so that the incident electric field can orient magnetic dipoles and incident magnetic field orients electric dipoles~\cite{Tellegen}. Physically this can be achieved by engineering the overlapping electric and magnetic dipole resonances of the meta-atom while properly breaking $\mathcal{P}$ and $\mathcal{T}$ symmetries.

By analogy, we suggest to compose $\psi$-metamaterial from the meta-atoms with overlapping and hybridized higher-order electric and magnetic multipole resonances. Excitation of higher-order multipoles affects the constitutive relations which take the following form~\cite{Achouri2021}
\begin{eqnarray}
    \mathbf{D} & = &  \mathbf{E}+4\pi\left(\mathbf{P}- \frac{1}{2}\nabla\cdot \overline{\overline{Q}} + \frac{1}{6} \nabla\cdot\nabla\cdot \overline{\overline{\overline{O}}}\right), \label{auxiliaryDdispersive}\\
    \mathbf{H} & = & \mathbf{B}-4\pi\left( \mathbf{M}-\frac{1}{2}\nabla \cdot \overline{\overline{S}} +\frac{1}{6}\nabla\cdot\nabla\cdot\overline{\overline{\overline{\Xi}}}\right), \label{auxiliaryHdispersive} 
\end{eqnarray}
where $\mathbf{P}$ and $\mathbf{M}$ are polarization and magnetization, respectively, $\overline{\overline{Q}}$ and $\overline{\overline{S}}$ are electric and magnetic quadrupole moments, $\overline{\overline{\overline{O}}}$ and $\overline{\overline{\overline{\Xi}}}$ are electric and magnetic octupole moments. 

In turn, higher-order multipole moments $\overline{\overline{Q}}$, $\overline{\overline{S}}$, $\overline{\overline{\overline{O}}}$ and $\overline{\overline{\overline{\Xi}}}$ are induced by the incident fields or their gradients. Inspecting Eqs.~\eqref{auxiliaryDdispersive}, \eqref{auxiliaryHdispersive}, we recover that the desired type of the constitutive relations can be obtained, in particular, if electric quadrupole moment $\overline{\overline{Q}}$ is induced by the gradient of magnetic field and magnetic quadrupole moment $\overline{\overline{S}}$ originates from the gradient of electric field.

The relevant polarizabilities governing the strength of such a response can be calculated using the perturbation theory and read~\cite{Supplement}
\begin{equation}
     s^e_{\alpha\beta\gamma \sigma} = q^m_{\gamma \sigma \alpha\beta} = \sum_j \mathcal{E}_{j n} \text{Re}\braket{n|S_{\alpha\beta}|j}\braket{j|Q_{\gamma\sigma}|n}\:.  \label{multipolemomentqq}
\end{equation}
Here $\ket{n}$ denote the eigenmodes of the particle, $Q_{\gamma\sigma}$ and $S_{\alpha\beta}$ are the operators of the respective components of electric and magnetic quadrupole moments, $\omega$ is the frequency of excitation, $\mathcal{E}_{j n} = (2/\hbar)\,\omega_{j n}/(\omega_{j n}^2-\omega^2)$ is the Lorentz factor, and $\omega_{jn}$ are the frequencies of the transitions between the different modes.

This expression, presented in a form resembling quantum-mechanical one, has two immediate implications. First, to maximize the desired response, electric and magnetic quadrupoles should resonante at the same or close frequencies. Second, the eigenmodes of the particle should be hybrid, containing the mixture of electric and magnetic quadrupoles thus rendering the two matrix elements simultaneously nonzero.

\begin{figure}[H]
    \centering
    \includegraphics{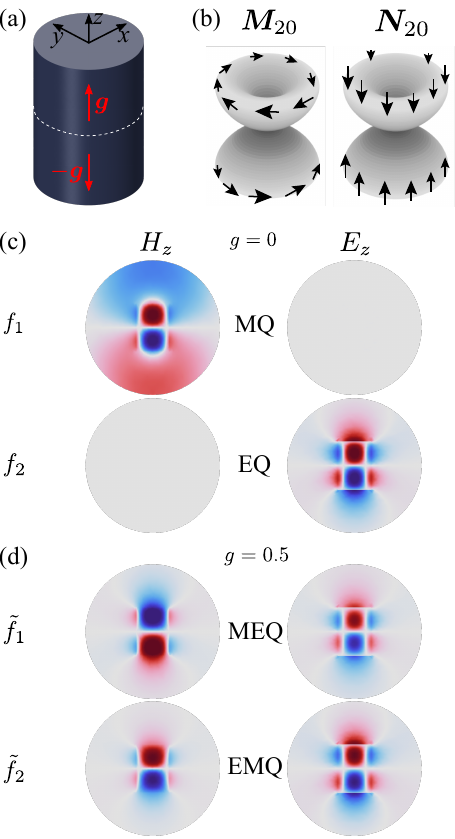}
    \caption{\label{fig:psi_meta-atom} Design of the meta-atom featuring $\psi$-response based on two oppositely magnetized ferrites stacked together. (a) Sketch of the proposed meta-atom with the following parameters: permittivity of both cylinders $\varepsilon = 14$, radius $r_0 = 8.2$~mm, total height $h_0 = 24$~mm. Red arrows show the direction of magnetization, white dashed line shows the boundary between the cylinders. (b) Schematic representation of the vector spherical harmonics corresponding to magnetic, $\bm{M}$, and electric, $\bm{N}$, quadrupoles ($l=2$) with $m=0$. Black arrows denote mutually orthogonal electric field distributions. (c) Field distributions corresponding to magnetic quadrupole (MQ, $H_z \ne 0,\ E_z = 0$) mode with eigenfrequency $f_1 = 4.59 - 0.01\,i$~GHz and electric quadrupole (EQ, $H_z = 0,\ E_z \ne 0$) mode with eigenfrequency $f_2 = 4.68 - 0.04\,i$~GHz when cylinders are not magnetized ($g = 0$). (d) Field distributions of the eigenmodes with frequencies $\tilde{f}_1 = 4.63 - 0.03\,i$ and $\tilde{f}_2 = 4.73 - 0.02\,i$~GHz in the presence of magnetization ($g = 0.5$). Both modes contain the mixture of electric and magnetic quadrupoles. The strength of the magneto-electric coupling is quantified by the complex ratio $a_{\mathrm{E}}(2,0)/a_{\mathrm{M}}(2,0)$, which is $0.38 + 0.04\,i$ and $-0.92 - 0.06\,i$ for the modes $\tilde{f}_1$ and $\tilde{f}_2$, respectively.}
\end{figure}

To meet those two requirements, we design the meta-atom depicted in Fig.~\ref{fig:psi_meta-atom}(a) consisting of two ferrite cylinders with the opposite directions of magnetization. Nonzero magnetization breaks $\mathcal{T}$ symmetry and renders the meta-atom non-reciprocal. In addition, both magnetization vectors change their sign upon mirror reflection in $Oxy$ plane which guarantees breaking of the $\mathcal{P}$ symmetry allowing the mixing of electric and magnetic eigenmodes~\cite{Poleva2023}. Note, however, that the combined $\mathcal{PT}$ symmetry is preserved for this design, which is exactly the symmetry properties needed for the effective axion as well as effective $\psi$-type response.




Each of the constituent cylinders is described in terms of the gyrotropic constitutive relations $\bm{B} = \mu \bm{H} + i\, \bm{H} \times \bm{g}$, where $\bm{g}$ is the gyration vector. In the case of $\bm{g} = \pm g \hat{\bm{z}}$, ferrite cylinders have the following antisymmetric permeability tensor:
\begin{equation}
    \hat{\mu} = 
    \begin{pmatrix}
        \mu & \pm i\,g & 0 \\
        \mp i\,g & \mu & 0 \\
        0 & 0 & 1
    \end{pmatrix}\:,
\end{equation}
where $+$ and $-$ signs correspond to upward and downward magnetization, respectively. To prove the feasibility of the $\psi$-type response, we perform full-wave numerical eigenmodes simulations with $\mu = 2$ and $g = 0.5$, that correspond to the realistic ferrites such as spinels in the frequency range of several GHz. Since we are interested only in a relatively narrow spectral range of quadrupole resonances, frequency dispersion of $\mu$ and $g$ is neglected.

Fields of electric and magnetic quadrupoles are defined by the vector spherical harmonics with indices $l=2$ and $m=-2, \dots ,2$~\cite{BohrenHuffman}. To avoid degeneracies, we focus on the modes with $m=0$ as depicted schematically in Fig.~\ref{fig:psi_meta-atom}(b).

As a first step, we simulate the non-magnetized version ($g = 0$) of the meta-atom. In this case, $\mathcal{P}$ and $\mathcal{T}$ symmetries are not broken and thus electric and magnetic quadrupole modes are perfectly decoupled. Their field profiles are shown in Fig.~\ref{fig:psi_meta-atom}(c). 

Next, we assume nonzero magnetization $g = 0.5$ and calculate the mode structure once again. Due to the breaking of time-reversal and spatial inversion symmetries, electric and magnetic modes mix, i.e. field distributions of the modes contain both electric and magnetic quadrupoles [Fig.~\ref{fig:psi_meta-atom}(d)]. To quantify the strength of mixing and relative phase shift, the ratio between the multipole coefficients $a_{\mathrm{M}}(2,0)$ and $a_{\mathrm{E}}(2,0)$ is calculated. Both modes experience significant mixing ($\mathrm{MQ}/\mathrm{EQ} \sim 0.4$ and $\mathrm{MQ}/\mathrm{EQ} \sim 1$, respectively) and the relative phase shifts between the electric and magnetic quadrupole moments are close to $0$ and $\pi$, respectively. Strong mixing of electric and magnetic quadrupole modes ensures that the array of such meta-atoms exhibits strong and resonant $\psi$-type response.

\textit{Discussion and conclusions.}~--- In summary, we have put forward a nonlocal generalization of axion electrodynamics featuring exotic electromagnetic properties. While the suggested material responds to the plane wave excitation in a way similar to the usual axion case, its response to the localized sources is profoundly different exhibiting vanishing Witten effect. While the respective constitutive relations may seem exotic from the first glance, we prove that they can be implemented experimentally using the platform of photonic metamaterials and optimizing the resonances of the constituent meta-atoms.

We believe that our findings open an exciting research avenue. From the fundamental perspective, of special interest might be the realization of dynamic $\psi$ fields and the possibility to couple them to dark matter axions using such metamaterials in axion search experiments. On the other hand, $\psi$-metamaterials can be of interest on their own opening routes in implementation of topologically protected photonic modes, nonreciprocal transmission of light on a chip,  nonreciprocal optical elements or wireless communication~\cite{Alireza2019,Guo2019,Yang2023}.

\begin{acknowledgments}
\textit{Acknowledgments.}~--- We acknowledge Maxim Mazanov for valuable discussions. Theoretical models were supported by the Russian Science Foundation (grant No.~23-72-10026), numerical simulations were supported by Priority 2030 Federal Academic Leadership Program. D.A.B. and M.A.G. acknowledge partial support by the Foundation
for the Advancement of Theoretical Physics and Mathematics ``Basis''.
\end{acknowledgments}


\bibliography{refs}

\end{document}


\title{Supplemental Materials: \\ Axion electrodynamics without Witten effect in metamaterials}

\author{Eduardo Barredo-Alamilla}
\email{eduardo.barredo@metalab.ifmo.ru}
\affiliation{School of Physics and Engineering, ITMO University, St. Petersburg, Russia}
\author{Daniel A. Bobylev}
\email{daniil.bobylev@metalab.ifmo.ru}
\affiliation{School of Physics and Engineering, ITMO University, St. Petersburg, Russia}
\author{Maxim A. Gorlach}
\email{m.gorlach@metalab.ifmo.ru}
\affiliation{School of Physics and Engineering, ITMO University, St. Petersburg, Russia}

\maketitle

\widetext

\setcounter{equation}{0}
\setcounter{figure}{0}
\setcounter{table}{0}
\setcounter{page}{1}
\setcounter{section}{0}
\makeatletter
\renewcommand{\theequation}{S\arabic{equation}}
\renewcommand{\thefigure}{S\arabic{figure}}
\renewcommand{\bibnumfmt}[1]{[S#1]}
\renewcommand{\citenumfont}[1]{S#1}

\tableofcontents

\section{Supplemental Note 1: Lagrangian formulation of $\psi$-electrodynamics}

Axion electrodynamics is an extension of Maxwell electrodynamics by the addition of a Chern-Simons term $\theta F_{\mu\nu}\tilde{F}^{\mu\nu}$ where $\theta$ is not a constant parameter but rather a pseudo-scalar field $\theta(\mathbf{x},t)$. For the $\psi$-electrodynamics we consider the modification of Maxwell theory by the addition of the term

\begin{equation}
    \mathcal{L}_\psi = -\frac{1}{16\pi} \psi F_{\mu\nu}\nabla^2\tilde{F}^{\mu\nu},
\end{equation}
to the Lagrangian density, where  $F_{\mu\nu}= \partial_\mu A_\nu - \partial_\nu A_\mu$ is the electromagnetic tensor, $\tilde{F}^{\mu\nu} = \frac{1}{2}\epsilon^{\mu\nu\alpha \beta}F_{\alpha\beta}$ is the dual electromagnetic tensor, $\epsilon^{\mu\nu\alpha \beta}$ is the Levi-Civita symbol with the convention $\epsilon^{0123}=1$, and $A^\mu=(\phi,\mathbf{A})$ is the gauge potential. The contraction between between the electromagnetic tensor and the Laplacian of the dual one is a total derivative

\begin{equation}
     F_{\mu\nu} \nabla^2 \tilde{F}^{\mu\nu} = 2 \partial_\mu \Big( A_\nu \nabla^2\tilde{F}^{\mu\nu} \Big).
\end{equation}
Then, the Lagrangian density can be manipulated taking into account that the total derivatives do not modify the physics:

 \begin{equation}
     \mathcal{L}= -\frac{1}{16\pi}F_{\mu\nu} F^{\mu\nu} -\frac{1}{c}J^\mu A_\mu + \frac{1}{8\pi} (\partial_\mu \psi)    (\nabla^2 \tilde{F}^{\mu\nu})  A_\nu,
 \end{equation}
where $J^\mu=(c\rho,\mathbf{J})$ is the conserved current. Written in this way, the Lagrangian has a form of a CPT-odd dimension-five extension when considering only spatial derivatives~\cite{Ferreira2019}. Since the Lagrangian involves third order derivatives of the gauge potential, Euler-Lagrange equation should be extended with the field derivatives of the same order

\begin{equation}
   \frac{\delta \mathcal{L}}{\delta A_\rho} - \partial_\sigma \left(  \frac{\delta \mathcal{L}}{\delta (\partial_\sigma A_\rho)}\right) + \partial_\lambda \partial_\sigma \left(  \frac{\delta \mathcal{L}}{\delta (\partial_\lambda \partial_\sigma A_\rho)}\right) - \partial_\gamma \partial_\lambda \partial_\sigma \left(  \frac{\delta \mathcal{L}}{\delta (\partial_\gamma \partial_\lambda \partial_\sigma A_\rho)}\right) = 0.
\end{equation}
The equations of motion of such Lagrangian read
\begin{equation}
    \partial_\mu F^{\mu\nu} = \frac{4\pi}{c}J^\nu - (\partial_\mu \psi) \nabla^2 \tilde{F}^{\mu\nu}.
\end{equation}
provided $\partial_\mu \psi $ is a constant. Under this assumption, we recover the equations of $\psi$-electrodynamics:
\begin{eqnarray}
    \nabla \cdot \mathbf{E} & = & 4\pi \rho -(\nabla \psi) \cdot (\nabla^2 \mathbf{B}), \\
    \nabla \times \mathbf{B} - \frac{1}{c}\frac{\partial\mathbf{E}}{\partial t}  & =  &\frac{4\pi}{c} \mathbf{J} + \frac{1}{c}\left(\frac{\partial \psi}{\partial t}\right) \mathbf{B} + (\nabla \psi)\times  \nabla^2\mathbf{E}.
\end{eqnarray}
Another pair of equation follows from the definition of the fields via potentials.

\section{Supplemental Note 2: Boundary conditions of $\psi$-electrodynamics}
\begin{figure}[h!]
	\centering
	\includegraphics[width=0.8\textwidth]{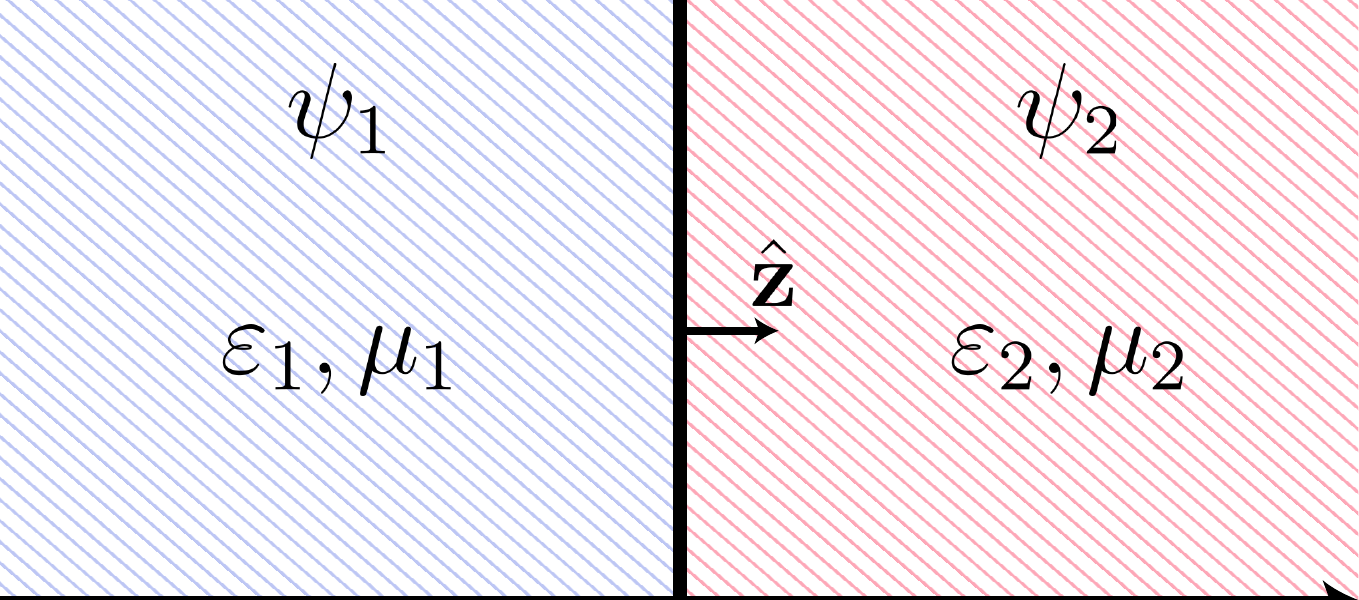}
	\caption{
		Geometry of the semi-infinite planar $\psi$-region. 
	}
	\label{FigS1BCplanar}
\end{figure}
In axion electrodynamics, modification to the electromagnetic response only arises when there is a gradient of the axion field. A similar behavior occurs in $\psi$-electrodynamics. In this section, we derive the boundary conditions for $\psi$-electrodynamics. Consider two semi-infinite metamaterials characterized by the different $\psi$, separated by a planar interface as shown in Fig.~\ref{FigS1BCplanar}. Since two metamaterials possess different constant values of $\psi$, we can write it as piecewise function

\begin{equation}
\psi(z) = 
\left\{
    \begin{array}{lr}
        \psi_1, & \text{if } z<0,\\
        \psi_2, & \text{if } z>0.
    \end{array}
\right.\label{piecewise_psiz}
\end{equation}
This can be expressed as $\psi(z) = H(z)\psi_2 + H(-z) \psi_1$, where $H(z)$ is the Heaviside function 
\begin{equation}
H(z) = 
\left\{
    \begin{array}{lr}
        1, & \text{if } z<0,\\
        0, & \text{if } z>0.
    \end{array}
\right.\label{heaviside}
\end{equation}
The gradient of $\psi(z)$ is 

\begin{equation}
    \nabla \psi(z) = (\psi_2 - \psi_1)\delta(z)\hat{\mathbf{z}}.
\end{equation}
Since the normal to the interface is $\hat{\mathbf{n}}=\hat{\mathbf{z}}$, the inhomogeneous modified Maxwell's equations read 

\begin{eqnarray} 
\nabla \cdot (\varepsilon\mathbf{E}) & = & 4\pi \rho+\tilde{\psi}\delta(z) (\nabla^2\mathbf{B}) \cdot \hat{\mathbf{z}}\label{psieqs1z},\\
\nabla \times \frac{\mathbf{B}}{\mu} -\frac{1}{c}\frac{\partial (\varepsilon \mathbf{E}) }{\partial t} & = & \frac{4\pi}{c} \mathbf{J} + \tilde{\psi}\delta(z) (\nabla^2\mathbf{E}) \times \hat{\mathbf{z}} \label{psieqs2z}.\end{eqnarray}
where $\tilde{\psi}=\psi_1-\psi_2$. We can regard these extra terms to Maxwell's equations as effective charges and current densities dependent of the electromagnetic fields that only exist on the interface and that satisfy the conservation of charge $ \nabla \cdot \mathbf{J}_\psi + \partial\rho_\psi /\partial t = 0$, 
\begin{equation}
     \rho_\psi = \frac{1}{4\pi} \tilde{\psi}\delta(z) (\nabla^2\mathbf{B}) \cdot \hat{\mathbf{z}}, \qquad \mathbf{J}_\psi = \frac{c}{4\pi} \tilde{\psi}\delta(z) (\nabla^2\mathbf{E}) \times \hat{\mathbf{z}}.
\end{equation}

\begin{figure}[t!]
	\centering
	\includegraphics[width=0.85\textwidth]{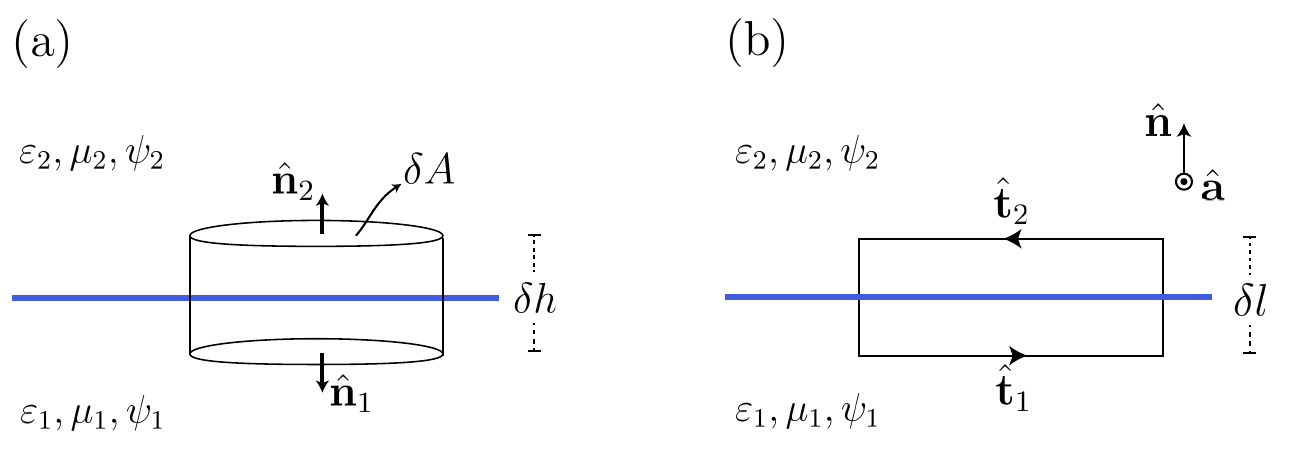}
	\caption{ a) Pillbox b) Thin rectangular loop.
	}
	\label{FigS2Pillbox}
\end{figure}

The boundary conditions can be obtained directly from Eqs.~(\ref{psieqs1z}) and (\ref{psieqs2z}). For simplicity, we assume that the external sources $\rho$, ${\bf j}$ are absent. First, we integrate Eq.~(\ref{psieqs1z}) over the volume of a pillbox of thickness $\delta h$ as shown in Fig.~\ref{FigS2Pillbox}(a)

\begin{equation}
\begin{split}
    \int_V dV \nabla \cdot(\varepsilon\mathbf{E}) =  &  \oint_S \varepsilon \mathbf{E} \cdot \hat{\mathbf{n}} dS = \varepsilon_1 \mathbf{E}_1 \cdot \hat{\mathbf{n}}_1 \delta A + \varepsilon_2 \mathbf{E}_2 \hat{\mathbf{n}}_2 \delta A = (\varepsilon_2 \mathbf{E}_2 - \varepsilon_1 \mathbf{E}_1)\cdot \hat{\mathbf{n}} \delta A \\
    = & \int_V dV \tilde{\psi}\delta(z) (\nabla^2 \mathbf{B})\cdot \hat{\mathbf{n}} = \int_{\partial V} \tilde{\psi} (\nabla^2 \mathbf{B})\cdot \hat{\mathbf{n}} dS |_{z=0} = \tilde{\psi} (\nabla^2 \mathbf{B}) \delta A |_{z=0}, 
\end{split}
\end{equation}
which yields the boundary condition 

\begin{equation}
    \varepsilon \mathbf{E}\cdot \hat{\mathbf{z}}|^{z=0^+}_{z=0^-} = \tilde{\psi} (\nabla^2\mathbf{B})\cdot \hat{\mathbf{z}}|_{z=0}
\label{BC1}\end{equation}

For the tangential components of the fields we consider a thin rectangular loop with the sides parallel to the interface as shown in Fig.~\ref{FigS2Pillbox}(b). We integrate Eq.~(\ref{psieqs2z}) over this open surface and consider $\delta l  \rightarrow 0$.

\begin{equation}
\begin{split}
    \int_S dS (\nabla \times \frac{\mathbf{B}}{\mu})\cdot \hat{\mathbf{a}} =  &  \oint_C \frac{\mathbf{B}}{\mu} \cdot d\mathbf{l} = \frac{\mathbf{B}_1}{\mu_1} \cdot\hat{\mathbf{t}}_1 \delta l + \frac{\mathbf{B}_2}{\mu_2} \cdot\hat{\mathbf{t}}_2 \delta l = \left( \frac{\mathbf{B}_2}{\mu_2} -\frac{\mathbf{B}_1}{\mu_1}\right) \cdot \hat{\mathbf{t}}_2 \delta l = -\hat{\mathbf{a}}\cdot \left[\left( \frac{\mathbf{B}_2}{\mu_2} -\frac{\mathbf{B}_1}{\mu_1}\right) \times \hat{\mathbf{n}}\right]  \delta l  \\
    = & \int_S dS (\tilde{\psi}\delta(z) (\nabla^2\mathbf{E}) \times \hat{\mathbf{z}})\cdot \hat{\mathbf{a}} = \int dl (\tilde{\psi}(\nabla^2 \mathbf{E})\times \hat{\mathbf{n}})\cdot \hat{\mathbf{a}}|_{z=0} = (\tilde{\psi}(\nabla^2 \mathbf{E})\times \hat{\mathbf{n}})\cdot \hat{\mathbf{a}} \delta l |_{z=0}, 
\end{split}
\end{equation}
where $\hat{\mathbf{a}}$ is a unitary vector pointing outside the page and $\hat{\mathbf{t}}_2 = \hat{\mathbf{b}} \times \hat{\mathbf{n}}$. We obtain the second boundary condition 

\begin{equation}
    \frac{\mathbf{B}}{\mu}\times  \hat{\mathbf{z}}|^{z=0^+}_{z=0^-} = -\tilde{\psi} (\nabla^2\mathbf{E})\times\hat{\mathbf{z}}|_{z=0} \label{BC2}
\end{equation}
The continuity equation for the homogeneous Maxwell equations are calculated straightforwardly,

\begin{equation}
      \mathbf{B}\cdot  \hat{\mathbf{z}}|^{z=0^+}_{z=0^-} = 0, \qquad  \mathbf{E}\times  \hat{\mathbf{z}}|^{z=0^+}_{z=0^-} = 0.
\end{equation}

As it is common in the theory of media with spatial dispersion, to make the description self-consistent, extra boundary conditions are needed. Specifically, one needs to ensure that the right-hand side of Eqs.~(\ref{BC1}) and (\ref{BC2}) is well-defined and these equations represent self-induced charge and current densities. To make the problem well-defined, we assume that the Laplacian of the electromagnetic fields at the interface can be written as the average of the Laplacian of the fields near the interface at both sides. Therefore, for vanishing free sources on the surface and $\mu$, $\varepsilon$ constants, the boundary conditions read

\begin{align}
   \varepsilon \mathbf{E}\cdot \hat{\mathbf{z}}|^{z=0^+}_{z=0^-} & = \tilde{\psi} (\nabla^2\mathbf{B})\cdot \hat{\mathbf{z}}|_{z=0}, & \frac{\mathbf{B}}{\mu}\times  \hat{\mathbf{z}}|^{z=0^+}_{z=0^-} & = -\tilde{\psi} (\nabla^2\mathbf{E})\times\hat{\mathbf{z}}|_{z=0}, \label{S1PsiBC1} \\
   \mathbf{B}\cdot  \hat{\mathbf{z}}|^{z=0^+}_{z=0^-} & = 0, & \mathbf{E}\times  \hat{\mathbf{z}}|^{z=0^+}_{z=0^-} & = 0, \label{S1PsiBC2}
\end{align}
where
\begin{eqnarray}
    (\nabla^2 \mathbf{B})\cdot \hat{\mathbf{z}} |_{z=0} & = & \frac{1}{2} \Big[ \nabla^2 \mathbf{B}|_{z=0^+} + \nabla^2 \mathbf{B}|_{z=0^-}  \Big] \cdot \hat{\mathbf{n}}, \label{LaplacianInter1} \\
     (\nabla^2 \mathbf{E})\times \hat{\mathbf{z}} |_{z=0} & = & \frac{1}{2} \Big[ \nabla^2 \mathbf{E}|_{z=0^+} + \nabla^2 \mathbf{E}|_{z=0^-}  \Big] \times \hat{\mathbf{n}}, \label{LaplacianInter2}
\end{eqnarray}
yielding the boundary conditions well-defined.

\section{Supplemental Note 3: Reflection and transmission matrices in $\psi$-electrodynamics}
\begin{figure}[t!]
	\centering
	\includegraphics[width=0.75\textwidth]{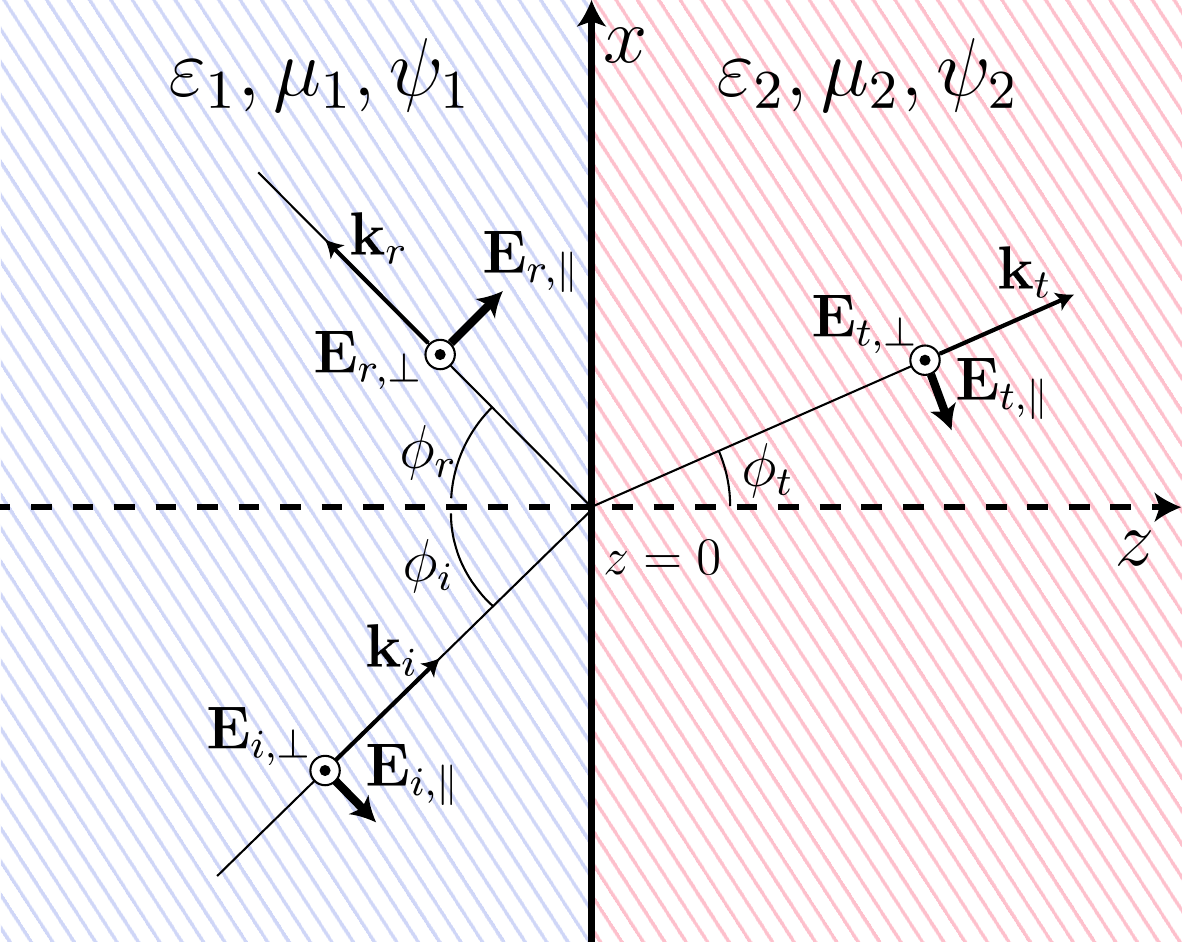}
	\caption{ Scattering of electromagnetic waves at a $\psi$-interface. The TE and TM has been decomposed.
	}
	\label{FigS3Scatter}
\end{figure}

In this section we derive the transmission and reflection matrix for a plane wave impinging at a $\psi$ interface. 
We consider the propagation of electromagnetic waves through the interface $\Sigma$ defined by the plane $z=0$. Suppose an electromagnetic wave propagates in the region $z<0$ and impinges on the interface at $z=0$ as illustrated in Fig.~\ref{FigS3Scatter}. Since Maxwell's equations do not change outside the interface, we assume that the plane electromagnetic wave has the form

\begin{equation}
    \begin{pmatrix} \mathbf{D}(t,\mathbf{r}) \\ \mathbf{H}(t,\mathbf{r})
    \end{pmatrix} = \begin{pmatrix} \mathbf{D}(\omega,\mathbf{k}) \\ \mathbf{H}(\omega,\mathbf{k})
    \end{pmatrix} e^{-i\omega t + i \mathbf{k}\cdot\mathbf{r}},
\end{equation}
and from the homogeneous Maxwell equations, one finds that 

\begin{equation}
    \mathbf{B} = \frac{1}{k_0} \mathbf{k}\times \mathbf{E}, \qquad \text{and} \qquad \mathbf{k}\cdot\mathbf{B} = 0, \label{HomEqPlaneWave}
\end{equation}
where $k_0=\omega/c$. Since there is a discontinuity in the fields at the interface given by Eq. (\ref{S1PsiBC1}) given by bound charges proportional to $\tilde{\psi}= \psi_1-\psi_2$, the propagation through the interface is  modified. The amplitudes of the reflected and transmitted wave can be obtained by applying the boundary conditions in Eq. (\ref{S1PsiBC1}). We are interested to see the difference in the transmission and reflection matrices. To do this, we use the subscripts $i$, $r$ and $t$ to tag the incident, reflected and transmitted wave

\begin{equation}
    \mathbf{E}_\alpha (t, \mathbf{r}) =  \mathbf{E}_\alpha (\omega, \mathbf{k})  e^{-i\omega t + i \mathbf{k}_\alpha\cdot\mathbf{r}}, \qquad \alpha = i, r, t.
\end{equation}

From now on, we consider the fields of the plane waves dependent only on frequency and wavevector. The boundary conditions implies that $\mathbf{k}_i\cdot\mathbf{r}=\mathbf{k}_r\cdot\mathbf{r}=\mathbf{k}_t\cdot\mathbf{r}$. 
We distinguish the electromagnetic fields perpendicular and parallel components with respect to the incidence plane

\begin{eqnarray}
    \mathbf{E}_\alpha & = &   \mathbf{E}_{\alpha,\perp}+\mathbf{E}_{\alpha,\parallel}  \\
    \mathbf{H}_\alpha & = & \mathbf{H}_{\alpha,\perp}+\mathbf{H}_{\alpha,\parallel}
\end{eqnarray}
where $\mathbf{E}_\parallel = (E^x_\parallel, 0, E^z_\parallel)$ and $\mathbf{E}_\perp = (0, E_\perp, 0)$.  
Using Eq. (\ref{HomEqPlaneWave}) we have

\begin{eqnarray}
   \mu k_0 \mathbf{H}_{\alpha,\parallel} & = & \mathbf{k}_\alpha \times \mathbf{E}_{\alpha,\perp} \label{HparallelEperp} \\
    \mu k_0 \mathbf{H}_{\alpha,\perp} & = & \mathbf{k}_\alpha \times \mathbf{E}_{\alpha,\parallel}
\end{eqnarray}
with the dispersion relation 
\begin{equation}
\mathbf{k}^2_\alpha = k_0^2\mu_\alpha\varepsilon_\alpha,
\label{dispersionrelation}\end{equation}
and $k_0 = \omega/c$. We start with the boundary conditions of the homogeneous Maxwell equations.
We can just consider the boundary condition of the electric field in Eq. (\ref{S1PsiBC2}) which yields

\begin{equation}
    \hat{\mathbf{n}} \times (\mathbf{E}^2-\mathbf{E}^1) = \hat{\mathbf{n}} \times (\mathbf{E}^2_\perp-\mathbf{E}^1_\perp) +  \hat{\mathbf{n}} \times  (\mathbf{E}^2_\parallel-\mathbf{E}^1_\parallel) = 0.\label{DiscomposedBC3}
\end{equation}
where $ \hat{\mathbf{n}} =  \hat{\mathbf{z}} $. 
Both perpendicular and parallel terms in Eq. (\ref{DiscomposedBC3}) are zero. This yields for the perpendicular components

\begin{equation}
    E_{i,\perp} + E_{r,\perp} - E_{t,\perp} = 0, \label{eqRandTmatrix1}
\end{equation}
which can also be deduced from the continuity of the normal components of magnetic field, Eq. (\ref{S1PsiBC2}). For the parallel components we have $\hat{\mathbf{n}}\times \mathbf{E}_\parallel = E^x_\parallel \hat{\mathbf{y}}$, hence 

\begin{equation}
    (E_{i,\parallel} - E_{r,\parallel})\cos\phi_i - E_{t,\parallel}\cos\phi_t = 0. 
\label{eqRandTmatrix2}\end{equation}
Because the electromagnetic fields are plane waves, the Laplacian of the fields at the interface,  Eqs. (\ref{LaplacianInter1}),(\ref{LaplacianInter2}),  yields

\begin{align}
      (\nabla^2\mathbf{B}) \cdot  \hat{\mathbf{z}}|_{z=0} & = -k_0^2\overline{\mu\varepsilon} \; \mathbf{B}\cdot \hat{\mathbf{z}}, & (\nabla^2\mathbf{E}) \times  \hat{\mathbf{z}}|_{z=0} & = -k_0^2\overline{\mu\varepsilon} \; \mathbf{E} \times  \hat{\mathbf{z}},
\label{averagesumdisprel}\end{align}
where $\overline{\mu\,\varepsilon} = \left(\mu_1\varepsilon_1 + \mu_2\varepsilon_2\right)/2$. More simply, we argue that in Eq. (\ref{S1PsiBC1}) the Laplacian operates as
\begin{equation}
\nabla^2 \xrightarrow{} = -k_0^2 \overline{\mu\,\varepsilon},
\end{equation} 
We use the boundary condition in Eq. (\ref{S1PsiBC1}) for the tangential components of the auxiliary field $\mathbf{H}$,

\begin{equation}
    \hat{\mathbf{n}} \times (\mathbf{H}^2_\parallel - \mathbf{H}^1_\parallel) = \tilde\psi\, k_0^2\, \overline{\mu\,\varepsilon}\,\hat{\mathbf{n}} \times \mathbf{E}_{t,\parallel}.
\label{PsiBC1tangential}\end{equation}
$\mathbf{H}_\parallel$ can be written in terms of the perpendicular components of the electric field by Eq. (\ref{HparallelEperp}) as follows

\begin{equation}
\hat{\mathbf{n}}\times\mathbf{H}_\parallel = \frac{1}{\mu k_0} \mathbf{n} \times (\mathbf{k}\times\mathbf{E}_\perp) = -\frac{1}{\mu k_0}  (\hat{\mathbf{n}}\cdot \mathbf{k})) \mathbf{E}_\perp,
\end{equation}
then we can rewrite Eq. (\ref{PsiBC1tangential}),

\begin{equation}
    \sqrt{\frac{\varepsilon_1}{\mu_1}}\left(E_{i,\perp}-E_{r,\perp}\right)\cos\phi_i - \sqrt{\frac{\varepsilon_1}{\mu_1}} E_{t,\perp} \cos\phi_t = -\tilde{\psi} \, k_0^2\, \overline{\mu\,\varepsilon}\, E_{t,\parallel} \label{eqRandTmatrix3} 
\end{equation}
Finally, we use the the boundary conditions for the normal components of the auxiliary field $\mathbf{D}$, in this case we also consider the parallel components of the electric field, 

\begin{equation}
    \hat{\mathbf{n}} \cdot (\mathbf{D}^2_\parallel-\mathbf{D}^1_\parallel) = - \tilde{\psi} \, k_0^2\, \overline{\mu\,\varepsilon}\, \hat{\mathbf{n}}\cdot\mathbf{B}_{t,\parallel} 
\end{equation}
We write $\mathbf{B}_{t,\parallel}$ in terms of the perpendicular components of the electric field,

\begin{equation}
  \hat{\mathbf{n}}  \cdot \mathbf{B}_{t,\parallel} = \frac{1}{k_0}\hat{\mathbf{n}} \cdot (\mathbf{k_t} \times \mathbf{E}_{t,\perp}) = \frac{1}{k_0} k_t \sin\phi_t E_{t,\perp}
\end{equation}
Therefore, the boundary condition for $\mathbf{D}$ yields,

\begin{equation}            \varepsilon_1(E^z_{i,\parallel}+E^z_{r,\parallel}) - \varepsilon_2 E^z_{t,\parallel} = \tilde{\psi} \, k_0^2\, \overline{\mu\,\varepsilon}\,\sqrt{\mu_1\varepsilon_1} \sin\phi_t E_{t,\perp}.
\end{equation}
By using $E^z_{\alpha,\parallel} = E_{\alpha,\parallel} \sin\phi_\alpha$ and the Snell's law we can write the last equation as

\begin{equation}
\sqrt{\frac{\varepsilon_1}{\mu_1}}\left(E_{i,\parallel}+E_{r,\parallel}\right) - \sqrt{\frac{\varepsilon_2}{\mu_2}}E_{t,\parallel} =  \tilde{\psi} \, k_0^2\, \overline{\mu\,\varepsilon}\,  E_{t,\perp}.
\label{eqRandTmatrix4}\end{equation}
This last equation can also be deducted from the perpendicular components of the auxiliary field $\mathbf{H}$ in Eq. (\ref{S1PsiBC1}). We have a set of four algebraic equations, Eq. (\ref{eqRandTmatrix1}), (\ref{eqRandTmatrix2}),  (\ref{eqRandTmatrix3}) and (\ref{eqRandTmatrix4}) to solve for four unknown quantities $E_{r,\perp}$, $E_{r,\parallel}$, $E_{t,\perp}$ and $E_{t,\parallel}$. First we write the $E_{r}$ components in terms of the $E_t$ and $E_i$ as

\begin{equation}
    E_{r,\parallel} =  E_{i,\parallel} -  E_{t,\parallel} \frac{\cos\phi_t}{\cos\phi_i}, \qquad  E_{r,\perp} = E_{t,\perp} - E_{i,\perp},\label{homogenousol1}
\end{equation}
and by substituting these two equations into Eq. (\ref{eqRandTmatrix3}) and Eq. (\ref{eqRandTmatrix4}) we obtain

\begin{eqnarray}
    \sqrt{\frac{\varepsilon_1}{\mu_1}}\left(2 E_{i,\parallel} -  E_{t,\parallel} \frac{\cos\phi_t}{\cos\phi_i}\right) - \sqrt{\frac{\varepsilon_2}{\mu_2}}E_{t,\parallel} & = &  \tilde{\psi} \, k_0^2\, \overline{\mu\,\varepsilon}\,  E_{t,\perp} \\
    \sqrt{\frac{\varepsilon_1}{\mu_1}}\left(2 E_{i,\perp}- E_{t,\perp}\right)\cos\phi_i - \sqrt{\frac{\varepsilon_1}{\mu_1}} E_{t,\perp} \cos\phi_t & = & -\tilde{\psi} \, k_0^2\, \overline{\mu\,\varepsilon}\, E_{t,\parallel}
\end{eqnarray}

After some algebra,

\begin{eqnarray}
    \left( \sqrt{\frac{\varepsilon_1}{\mu_1}}\cos\phi_t+\sqrt{\frac{\varepsilon_2}{\mu_2}}\cos\phi_i \right)E_{t,\parallel} +\tilde{\psi} \, k_0^2\, \overline{\mu\,\varepsilon}\, \cos\phi_i E_{t,\perp}& = & 2\sqrt{\frac{\varepsilon_1}{\mu_1}}\cos\phi_i E_{i,\parallel}  \\
    \left( \sqrt{\frac{\varepsilon_1}{\mu_1}}\cos\phi_i+\sqrt{\frac{\varepsilon_2}{\mu_2}}\cos\phi_t \right)E_{t,\perp} -\tilde{\psi} \, k_0^2\, \overline{\mu\,\varepsilon}\, \cos\phi_t E_{t,\parallel}& = & 2\sqrt{\frac{\varepsilon_1}{\mu_1}}\cos\phi_i E_{i,\perp},
\end{eqnarray}
which can be further express in matrix form

\begin{equation}
    \begin{pmatrix}
        \sqrt{\frac{\varepsilon_1}{\mu_1}}\cos\phi_t+\sqrt{\frac{\varepsilon_2}{\mu_2}}\cos\phi_i & \tilde{\psi} \, k_0^2\, \overline{\mu\,\varepsilon}\, \cos\phi_i  \\
        -\tilde{\psi} \, k_0^2\, \overline{\mu\,\varepsilon}\, \cos\phi_t & \sqrt{\frac{\varepsilon_1}{\mu_1}}\cos\phi_i+\sqrt{\frac{\varepsilon_2}{\mu_2}}\cos\phi_t 
    \end{pmatrix} 
    \begin{pmatrix}
        \vphantom{\sqrt{\frac{\varepsilon_1}{\mu_1}}} E_{t,\parallel}\\ \vphantom{\sqrt{\frac{\varepsilon_1}{\mu_1}}} E_{t,\perp} 
    \end{pmatrix} = 2\sqrt{\frac{\varepsilon_1}{\mu_1}}\cos\phi_i \begin{pmatrix}
        \vphantom{\sqrt{\frac{\varepsilon_1}{\mu_1}}}1 & 0  \\
        0 & \vphantom{\sqrt{\frac{\varepsilon_1}{\mu_1}}}1 
    \end{pmatrix}
    \begin{pmatrix}
        \vphantom{\sqrt{\frac{\varepsilon_1}{\mu_1}}} E_{i,\parallel}\\ \vphantom{\sqrt{\frac{\varepsilon_1}{\mu_1}}} E_{i,\perp} 
    \end{pmatrix}
\label{matrixtrans1}\end{equation}
We solve this matrix equation for the  components of the transmitted electric field. First we calculate the inverse matrix of the left-hand side of Eq. (\ref{matrixtrans1}),
\begin{equation}
     \begin{pmatrix}
        \sqrt{\frac{\varepsilon_1}{\mu_1}}\cos\phi_t+\sqrt{\frac{\varepsilon_2}{\mu_2}}\cos\phi_i & \tilde{\psi} \, k_0^2\, \overline{\mu\,\varepsilon}\, \cos\phi_i  \\
        -\tilde{\psi} \, k_0^2\, \overline{\mu\,\varepsilon}\, \cos\phi_t & \sqrt{\frac{\varepsilon_1}{\mu_1}}\cos\phi_i+\sqrt{\frac{\varepsilon_2}{\mu_2}}\cos\phi_t 
    \end{pmatrix}^{-1} = \frac{1}{\mathcal{D(\tilde{\psi})}} 
    \begin{pmatrix}
        \sqrt{\frac{\varepsilon_1}{\mu_1}}\cos\phi_i+\sqrt{\frac{\varepsilon_2}{\mu_2}}\cos\phi_t & -\tilde{\psi} \, k_0^2\, \overline{\mu\,\varepsilon}\, \cos\phi_i  \\
        \tilde{\psi} \, k_0^2\, \overline{\mu\,\varepsilon}\, \cos\phi_t & \sqrt{\frac{\varepsilon_1}{\mu_1}}\cos\phi_t+\sqrt{\frac{\varepsilon_2}{\mu_2}}\cos\phi_i 
    \end{pmatrix} \label{inversemat1}
\end{equation}
where $\mathcal{\tilde{\psi}}$ is the determinant of the matrix
\begin{equation}
   \mathcal{D}(\mathcal{\Tilde{\psi}}) = \cos\phi_i\cos\phi_t\left( \frac{\varepsilon_1}{\mu_1}+\frac{\varepsilon_2}{\mu_2} +\sqrt{\frac{\varepsilon_1\varepsilon_2}{\mu_1\mu_2}}\left(\frac{\cos\phi_i}{\cos\phi_t}+\frac{\cos\phi_t}{\cos\phi_i}\right)+ \Tilde{\psi}^2\, k_0^4\, \overline{\mu\,\varepsilon}^2\,\right)= \cos\phi_i\cos\phi_t\Delta(\tilde{\psi}),
\end{equation}
and then we multiply Eq. (\ref{matrixtrans1}) by the left with the inverse in Eq. (\ref{inversemat1}) yielding

\begin{equation}
    \begin{pmatrix}
        \vphantom{\sqrt{\frac{\varepsilon_1}{\mu_1}}} E_{t,\parallel}\\ \vphantom{\sqrt{\frac{\varepsilon_1}{\mu_1}}} E_{t,\perp} 
    \end{pmatrix} = \frac{2\sqrt{\frac{\varepsilon_1}{\mu_1}}\sec\phi_t}{\Delta(\tilde{\psi})}
    \begin{pmatrix}
        \sqrt{\frac{\varepsilon_1}{\mu_1}}\cos\phi_i+\sqrt{\frac{\varepsilon_2}{\mu_2}}\cos\phi_t & -\tilde{\psi} \, k_0^2\, \overline{\mu\,\varepsilon}\, \cos\phi_i  \\
        \tilde{\psi} \, k_0^2\, \overline{\mu\,\varepsilon}\, \cos\phi_t & \sqrt{\frac{\varepsilon_1}{\mu_1}}\cos\phi_t+\sqrt{\frac{\varepsilon_2}{\mu_2}}\cos\phi_i 
    \end{pmatrix} 
    \begin{pmatrix}
        \vphantom{\sqrt{\frac{\varepsilon_1}{\mu_1}}} E_{i,\parallel}\\ \vphantom{\sqrt{\frac{\varepsilon_1}{\mu_1}}} E_{i,\perp} 
    \end{pmatrix}
\label{transmatrix1}\end{equation}
Now, we solve reflected components of the electric field by using the Eq. (\ref{homogenousol1}) which yields

\begin{equation}
    \begin{pmatrix}
        \vphantom{\sqrt{\frac{\varepsilon_1}{\mu_1}}} E_{r,\parallel}\\ \vphantom{\sqrt{\frac{\varepsilon_1}{\mu_1}}} E_{r,\perp} 
    \end{pmatrix} = \frac{1}{\Delta(\tilde{\psi})}\begin{pmatrix}
        \left(\frac{\varepsilon_2}{\mu_2}- \frac{\varepsilon_1}{\mu_1}+ \Tilde{\psi}^2\, k_0^4\, \overline{\mu\,\varepsilon}^2\, \right) +\sqrt{\frac{\varepsilon_1\varepsilon_2}{\mu_1\mu_2}}\chi_- & 2 \sqrt{\frac{\varepsilon_1}{\mu_1}}\tilde{\psi} \, k_0^2\, \overline{\mu\,\varepsilon}\, \\
        2 \sqrt{\frac{\varepsilon_1}{\mu_1}}\tilde{\psi} \, k_0^2\, \overline{\mu\,\varepsilon}\, & -\left(\frac{\varepsilon_2}{\mu_2}- \frac{\varepsilon_1}{\mu_1}+ \Tilde{\psi}^2\, k_0^4\, \overline{\mu\,\varepsilon}^2\, \right) +\sqrt{\frac{\varepsilon_1\varepsilon_2}{\mu_1\mu_2}}\chi_-
    \end{pmatrix} 
    \begin{pmatrix}
        \vphantom{\sqrt{\frac{\varepsilon_1}{\mu_1}}} E_{i,\parallel}\\ \vphantom{\sqrt{\frac{\varepsilon_1}{\mu_1}}} E_{i,\perp} 
    \end{pmatrix}
\label{reflmatrix1}\end{equation}
where
\begin{equation}
    \Delta(\tilde{\psi}) = \left( \frac{\varepsilon_1}{\mu_1}+\frac{\varepsilon_2}{\mu_2}+ \Tilde{\psi}^2\, k_0^4\, \overline{\mu\,\varepsilon}^2\, +\sqrt{\frac{\varepsilon_1\varepsilon_2}{\mu_1\mu_2}}\chi_+ \right),
\end{equation}
and
\begin{equation}
    \chi_\pm = \frac{\cos\phi_i}{\cos\phi_t} \pm \frac{\cos\phi_t}{\cos\phi_i}.
\end{equation}
Finally we obtain the transmission  and reflection matrices for the $\psi$-interface

\begin{eqnarray}
\bm{T} & = & \frac{2\sqrt{\frac{\varepsilon_1}{\mu_1}}\sec\phi_t}{\Delta(\tilde{\psi})} \begin{pmatrix}
        \sqrt{\frac{\varepsilon_1}{\mu_1}}\cos\phi_i+\sqrt{\frac{\varepsilon_2}{\mu_2}}\cos\phi_t & -\tilde{\psi} \, k_0^2\, \overline{\mu\,\varepsilon}\, \cos\phi_i  \\
        \tilde{\psi} \, k_0^2\, \overline{\mu\,\varepsilon}\, \cos\phi_t & \sqrt{\frac{\varepsilon_1}{\mu_1}}\cos\phi_t+\sqrt{\frac{\varepsilon_2}{\mu_2}}\cos\phi_i 
    \end{pmatrix}  \\
\bm{R} & = & \frac{1}{\Delta(\tilde{\psi})}\begin{pmatrix}
        \left(\frac{\varepsilon_2}{\mu_2}- \frac{\varepsilon_1}{\mu_1}+ \Tilde{\psi}^2\, k_0^4\, \overline{\mu\,\varepsilon}^2\, \right) +\sqrt{\frac{\varepsilon_1\varepsilon_2}{\mu_1\mu_2}}\chi_- & 2 \sqrt{\frac{\varepsilon_1}{\mu_1}}\tilde{\psi} \, k_0^2\, \overline{\mu\,\varepsilon}\, \\
        2 \sqrt{\frac{\varepsilon_1}{\mu_1}}\tilde{\psi} \, k_0^2\, \overline{\mu\,\varepsilon}\, & -\left(\frac{\varepsilon_2}{\mu_2}- \frac{\varepsilon_1}{\mu_1}+ \Tilde{\psi}^2\, k_0^4\, \overline{\mu\,\varepsilon}^2\, \right) +\sqrt{\frac{\varepsilon_1\varepsilon_2}{\mu_1\mu_2}}\chi_-
    \end{pmatrix} 
\end{eqnarray}

\section{Supplemental Note 4: Surface waves in $\psi$-electrodynamics}
In this section we present the derivation of the dispersion relation of surface waves that propagates along the $\psi$-interface. Consider the case when a surface wave propagates just along the $x$-axis, then, due to the translational symmetry along the $y$-axis there is no dependence of the $y$ variable. We assume the solutions for the surface wave of the form 

\begin{equation}
    \begin{pmatrix} \mathbf{D}(t,\mathbf{r}) \\ \mathbf{H}(t,\mathbf{r})
    \end{pmatrix} = \begin{pmatrix} \mathbf{D}_0(\omega,\mathbf{k}) \\ \mathbf{H}_0(\omega,\mathbf{k})
    \end{pmatrix} e^{-i\omega t + i \beta x}.\label{ansatzsurfacewaves}
\end{equation}
where $\beta$ is the propagation constant. The wave equations are as usual

\begin{equation}
    \nabla^2 \mathbf{E} -\frac{\mu}{c^2}\frac{\partial^2}{\partial t^2}\mathbf{D}=0, \qquad \nabla^2 \mathbf{H} -\frac{\varepsilon}{c^2}\frac{\partial^2}{\partial t^2}\mathbf{B}=0,
\end{equation}
The wave equation for the electric fields yields,

\begin{equation}
    \frac{\partial^2 \mathbf{E}_0}{\partial x^2} + \frac{\partial^2 \mathbf{E}_0}{\partial z^2} + k_0^2 \mu \mathbf{D}_0 = 0,
\label{convwaveeq}\end{equation}
because there is no dependence on the $y$ variable. By substituting the ansatz in Eq. (\ref{ansatzsurfacewaves}) into the Faraday law, $\nabla \times \mathbf{E} = ik_0 \mathbf{B}$ and $\nabla \times \mathbf{H} = -i k_0 \mathbf{D}$, we get

\begin{align}
B_{0x}   & = \frac{i}{k_0}\frac{\partial E_{0y}}{\partial z}, &  B_{0y}   & = \frac{\beta}{k_0}\frac{\partial E_{0y}}{\partial z}, & i k_0 B_{0z}   & = \frac{\partial E_{0x}}{\partial z} - i \beta E_{0z}, \\ 
D_{0x}   & = - \frac{i}{k_0}\frac{\partial H_{0y}}{\partial z}, &  D_{0y}   & = -\frac{\beta}{k_0}\frac{\partial E_{0y}}{\partial z}, & i k_0 D_{0z}   & = - \frac{\partial H_{0x}}{\partial z} + i\beta H_{0z}.
\end{align}
We can solve this set of equations in terms of the components $E_{0x}$ and $E_{0y}$. The wave equations to solve, Eq. (\ref{convwaveeq}), are

\begin{align}
      \frac{\partial^2 E_{0y}}{\partial z^2} + (k_0^2 \mu \varepsilon -\beta^2)E_{0y} & = 0, &   \frac{\partial^2 E_{0x}}{\partial z^2} + (k_0^2 \mu \varepsilon -\beta^2)E_{0x} & = 0,
\end{align}
and we expect the solutions of these fields to decrease at $|z| \to \infty$. Hence we need the propagation constant, $\beta$, such as

\begin{align}
    q^2 & = \beta^2 -k_0^2 \mu_1\varepsilon_1 > 0, & p^2 & =  \beta^2 -k_0^2 \mu_2\varepsilon_2 > 0.
\end{align}
At $z<0$
\begin{align}
E^{(1)}_{0x}   & = A e^{qz}, &  E^{(1)}_{0y}   & = B e^{qz}, &  E^{(1)}_{0z}   & = \frac{\beta}{i q} A e^{qz}, \\ 
H^{(1)}_{0x}   & = i \frac{q}{k_0 \mu_1} B e^{qz},  &  H^{(1)}_{0y}   & = i \frac{k_0 \varepsilon_1}{ q} A e^{qz} &  H^{(1)}_{0z}   & = \frac{\beta}{k_0 \mu_1}B e^{qz}.
\end{align}
At $z>0$
\begin{align}
E^{(2)}_{0x}   & = C e^{-pz}, &  E^{(2)}_{0y}   & = D e^{-pz}, &  E^{(2)}_{0z}   & = \frac{i \beta}{ p} C e^{-pz}, \\ 
H^{(2)}_{0x}   & = - i \frac{p}{k_0 \mu_2} D e^{-pz},  &  H^{(2)}_{0y}   & = \frac{k_0 \varepsilon_2}{i p} C e^{-pz}, &  H^{(2)}_{0z}   & = \frac{\beta}{k_0 \mu_2}D e^{-pz}.
\end{align}
In the following we use the boundary conditions, Eqs. (\ref{S1PsiBC1})-(\ref{S1PsiBC2}), to obtain the relationship between the amplitudes of the fields on either side of the interface and the dispersion equation of the surface wave propagating along the interface. First, from the continuity of the transversal electric field, we obtain  $E^{(1)}_{0x} = E^{(2)}_{0x}$ and $E^{(1)}_{0y} = E^{(2)}_{0y}$ at $z=0$, which yields $A = C$ and $B=D$. For the Laplacian of the electromagnetic fields at the interface, we consider that it can be written as the average sum of the dispersion relation of each side of the interface as in Eq. (\ref{averagesumdisprel}). Hence

\begin{align}
      \hat{\mathbf{z}} \cdot (\nabla^2\mathbf{B})   |_{z=0} & = - \, k_0^2\, \overline{\mu\,\varepsilon}\, \frac{\beta}{k_0} B, & \hat{\mathbf{z}} \times (\nabla^2\mathbf{E}) |_{z=0} & = -\, k_0^2\, \overline{\mu\,\varepsilon}\, \; (B, -A).
\label{S3averagesumdisprel}\end{align}
Then, by applying the boundary conditions in Eq. (\ref{S1PsiBC1}) we have

\begin{align}
\left( \frac{\varepsilon_2}{p}+\frac{\varepsilon_1}{q}\right) A & = i \tilde{\psi}\, k_0\, \overline{\mu\,\varepsilon} B, & \left( \frac{p}{\mu_2}+\frac{q}{\mu_1}\right) B & = i \tilde{\psi}\, k_0^3\, \overline{\mu\,\varepsilon} A.
\end{align}

This finally yields the dispersion relation of the form

\begin{equation}
\left( \frac{p}{\mu_2}+\frac{q}{\mu_1}\right)\left( \frac{p}{\varepsilon_2}+\frac{q}{\varepsilon_1}\right) + \Tilde{\psi}^2\,k_0^4 \,\overline{\mu\,\varepsilon}^2 \,\frac{pq}{\varepsilon_1\varepsilon_2}=0.
\end{equation}

\section{Supplemental Note 5: Spherical $\psi$-metamaterial in constant electric field}

Consider a sphere made of $\psi$-metamaterial of radius $R$ with parameter $\psi$ and $\varepsilon=\mu=1$ in a constant electric field that points along the z direction. In spherical coordinates the potential associated to the constant field is 
\begin{equation}
    \phi(z) = -E_0 z = -E_0 r \cos\theta = -E_0 P_1(\cos\theta),
\end{equation}
where $P_1$ is the first Legendre polynomial. The applied field only contributes to $l=1$ terms. Since there are no sources inside and outside of the sphere, one can use the notion of electric and magnetic potentials and present the solutions to the Laplace equations in the form 

\begin{eqnarray}
    \phi^{\text{out}}_E & = & -E_0 r P_1 (\cos\theta) + \sum_{l=0} \frac{A_l}{r^{l+1}} P_l (\cos\theta) \\
    \phi^{\text{out}}_M & = &  \sum_{l=0} \frac{B_l}{r^{l+1}} P_l (\cos\theta)
\end{eqnarray}
for the outer fields, and

\begin{eqnarray}
    \phi^{\text{in}}_E & = & \sum_{l=0} C_l r^l P_l (\cos\theta) \\
    \phi^{\text{in}}_M & = &  \sum_{l=0} D_l r^l P_l (\cos\theta)
\end{eqnarray}
for the inner fields. The boundary conditions, Eqs. (\ref{S1PsiBC1}-\ref{S1PsiBC2}), for the electric and magnetic potential become 

\begin{align}
   \left. \left[\frac{\partial \phi^{\text{out}}_E}{\partial_r}-\frac{\partial\phi^{\text{in}}_E}{\partial_r}\right]\right|_{r=R} & = \left. \tilde{\psi} \frac{\partial}{\partial r} (\nabla^2 \phi_M)\right|_{r=R}, & \left. \left[\frac{\partial \phi^{\text{out}}_M}{\partial_\theta}-\frac{\partial\phi^{\text{in}}_M}{\partial_\theta}\right]\right|_{r=R} & = -\left. \tilde{\psi} \frac{\partial}{\partial r} (\nabla^2 \phi_E)\right|_{r=R}, \label{S4PsiBC1} \\
    \left. \left[\frac{\partial \phi^{\text{out}}_M}{\partial_r}-\frac{\partial\phi^{\text{in}}_M}{\partial_r}\right]\right|_{r=R} &  = 0, & \left. \left[\frac{\partial \phi^{\text{out}}_E}{\partial_\theta}-\frac{\partial\phi^{\text{in}}_E}{\partial_\theta}\right]\right|_{r=R} & = 0, \label{S4PsiBC2}
\end{align}
where we have used $\hat{\mathbf{n}}\cdot \nabla^2 (\nabla\phi) = \hat{\mathbf{n}}\cdot \nabla (\nabla^2\phi) = \partial/\partial_r(\nabla^2\phi)$ and $\hat{\mathbf{n}}\times \nabla^2 (\nabla\phi) = 1/r \partial/\partial_\theta(\nabla^2\phi)\hat{\boldsymbol{\varphi}}$. Now, since the scalar and magnetic potentials are solutions of the Laplace equation, $\nabla^2\phi_{E,M}=0$ in both sides of the interface, and since the bound charges at the interface depend on these terms, we obtain 

\begin{align}
   \left. \left[\frac{\partial \phi^{\text{out}}_E}{\partial_r}-\frac{\partial\phi^{\text{in}}_E}{\partial_r}\right]\right|_{r=R} & =0, & \left. \left[\frac{\partial \phi^{\text{out}}_M}{\partial_\theta}-\frac{\partial\phi^{\text{in}}_M}{\partial_\theta}\right]\right|_{r=R} & = 0, \label{S4PsiBC11} \\
    \left. \left[\frac{\partial \phi^{\text{out}}_M}{\partial_r}-\frac{\partial\phi^{\text{in}}_M}{\partial_r}\right]\right|_{r=R} &  = 0, & \left. \left[\frac{\partial \phi^{\text{out}}_E}{\partial_\theta}-\frac{\partial\phi^{\text{in}}_E}{\partial_\theta}\right]\right|_{r=R} & = 0, \label{S4PsiBC22}
\end{align}
which yield that the potentials take the form

\begin{align}
   \phi^{\text{out}}_E & = -E_0 r P_1(\cos\theta), & \phi^{\text{out}}_M = 0, \label{S4outpotential} \\
    \phi^{\text{in}}_E & = -E_0 r P_1(\cos\theta), & \phi^{\text{in}}_M = 0. \label{S4inpotential}
\end{align}
This implies that under a constant field, the $\psi$-sphere is transparent. Unlike the axion sphere in which magnetic dipoles are formed. We can argue that the analogous effect will happen in the case of an applied constant magnetic field.

\section{Supplemental Note 6: Spherical $\psi$-metamaterial with point dipole}
Consider now the same $\psi$-sphere surrounding an electric dipole. In spherical coordinates the potential associated to the field of a point dipole in the origin is
\begin{equation}
    \phi(z) = \frac{\mathbf{p}_0\cdot \hat{\mathbf{r}}}{r^2}=\frac{p_0 \cos\theta}{r^2}= \frac{p_0}{r^2}P_1(\cos\theta),
\end{equation}
where $\mathbf{p}_0$ is the dipole moment. Note that the respective potential is a solution to the Laplace equation. For the same reason as in the case of a $\psi$-metamaterial in a constant electric field, the Laplacian of the electric and magnetic potentials will be zero, yielding the same boundary conditions as in Eqs. (\ref{S4PsiBC11}) and (\ref{S4PsiBC22}), which in turn yields the potentials
\begin{align}
   \phi^{\text{out}}_E & = \frac{p_0}{r^2} P_1(\cos\theta), & \phi^{\text{out}}_M = 0, \label{S5outpotential} \\
    \phi^{\text{in}}_E & = \frac{p_0}{r^2} P_1(\cos\theta), & \phi^{\text{in}}_M = 0, \label{S5inpotential}
\end{align}
Therefore, a dipole inside a $\psi$-sphere does not induce magnetic and electric dipoles unlike in the axion sphere where magnetic dipoles and constants fields are induced. The $\psi$-sphere is transparent under static sources.

\section{Supplemental Note 7: Theory of multipoles }

Below, we provide the derivation of the multipolar expansion up to the octupolar responses following the approach of Ref.~\cite{Achouri2021}. For consistency with Ref.~\cite{Achouri2021}, we exploit SI systems of units in this section.
It is known that the electromagnetic responses of a metamaterial are expressed in terms of the constitutive relations. We will consider $\mathbf{D}$ and $\mathbf{H}$ in terms of $\mathbf{E}$ and $\mathbf{B}$. First, we expand the electric current density $\mathbf{J}$ in terms of multipoles from the vector potential $\mathbf{A}(\mathbf{x})$ in SI units, 

\begin{equation}
    \mathbf{A}(\mathbf{x}) =\frac{\mu_0}{4\pi} \int \mathbf{J}(\mathbf{x}) \frac{e^{ik|\mathbf{x}-\mathbf{x}'|}}{|\mathbf{x}-\mathbf{x}'|^3}d^3 x',
\label{multipolepotvec}\end{equation}
where $k$ is the wavevector and we have assumed the sources to vary in time as $e^{-i\omega t}$. By doing a Taylor expansion on the free Green's function we obtain

\begin{equation}
    \frac{e^{ik|\mathbf{x}-\mathbf{x}'|}}{|\mathbf{x}-\mathbf{x}'|^3} = \sum_{n=0}^\infty \frac{1}{n!} (-\mathbf{r}'\cdot \nabla)^n \frac{e^{ikr}}{r}.
\label{greenfuncexpan}\end{equation}
Substituting  Eq. (\ref{greenfuncexpan}) into Eq. (\ref{multipolepotvec}), yields

\begin{equation}
\begin{split}
    \mathbf{A}(\mathbf{x}) & = \frac{\mu_0}{4\pi} \left[ \int \mathbf{J}(\mathbf{x})d^3 x' - \left(\int \mathbf{J}(\mathbf{x}) (\mathbf{r}' \cdot \nabla) d^3 x'\right) +\frac{1}{2}\left(\int \mathbf{J}(\mathbf{x}) (\mathbf{r}'\cdot\nabla)(\mathbf{r}'\cdot\nabla) d^3 x'\right)\right. \\
    & \left.+ \frac{1}{6}\left(\int \mathbf{J}(\mathbf{x}) (\mathbf{r}'\cdot\nabla)(\mathbf{r}'\cdot\nabla)(\mathbf{r}'\cdot\nabla) d^3 x'\right) +\cdots \right] \frac{e^{ikr}}{r},
\end{split}
\end{equation}
where the first term is related to the electric dipole, the second term to the magnetic dipole and electric quadrupole, the third one to the magnetic quadrupole and electric octupole, and the fourth to the magnetic octupole and the electric hexadecapole. Because of the current conservation and given the sinusoidally time-varying currents $e^{-i\omega t}$,

\begin{equation}
    \nabla \cdot J(\mathbf{r}) = i \omega \rho (\mathbf{r}),
\end{equation}
and then, the integrals like

\begin{equation}
    \int \Psi\,\mathbf{r'} \rho(\mathbf{r}') dV' = -\frac{i}{\omega} \int \Psi\,\mathbf{r'} \nabla'\cdot \mathbf{J}(\mathbf{r}') dV' = \frac{i}{\omega} \int \mathbf{J}(\mathbf{r}') \cdot \nabla' \Psi(\mathbf{r}') dV'.
\end{equation}
Therefore, we can express the multipoles moments in terms of the currents $\mathbf{J}$. For the electric multipoles we have
\begin{eqnarray}
    p_\alpha & = & \int \rho(\mathbf{r}') x_\alpha' dV' = \frac{i}{\omega}\int J_\alpha(\mathbf{r}') dV', \\
    q_{\alpha\beta} & = & \int \rho(\mathbf{r}') x_\alpha'x_\beta' dV' = \frac{i}{\omega}\int (J_\alpha x'_\beta + J_\beta x'_\alpha ) dV', \\
     o_{\alpha\beta\gamma} & = & \int \rho(\mathbf{r}') x_\alpha'x_\beta' x_\gamma'dV' = \frac{i}{\omega}\int (J_\alpha x'_\beta x'_\gamma + J_\beta x'_\alpha x'_\gamma + J_\gamma x'_\alpha x'_\beta) dV',
\end{eqnarray}
and for the magnetic multipoles
\begin{eqnarray}
    m_\alpha & = & \frac{1}{2}\int (\mathbf{r}' \times \mathbf{J})_\alpha dV'  \\
    s_{\alpha\beta} & = & \frac{2}{3} \int (\mathbf{r}' \times \mathbf{J})_\beta x'_\alpha dV' , \\
     \xi_{\alpha\beta\gamma} & =  & \frac{3}{4} \int (\mathbf{r}' \times \mathbf{J})_\gamma x'_\alpha x'_\beta dV'.
\end{eqnarray}
We are interested in the case of a bulk homogeneous finite medium. After been excited by electromagnetic fields, it exhibits a given distribution of current density. In terms of the current density the multipole moments are
\begin{align}
     P_\alpha & =  \frac{i}{\omega} J_\alpha(\mathbf{r}'), & Q_{\alpha\beta} & =   \frac{i}{\omega} (J_\alpha x'_\beta + J_\beta x'_\alpha ), &  O_{\alpha\beta\gamma} & =  \frac{i}{\omega} (J_\alpha x'_\beta x'_\gamma + J_\beta x'_\alpha x'_\gamma + J_\gamma x'_\alpha x'_\beta), \\
      M_\alpha & = \frac{1}{2} (\mathbf{r}' \times \mathbf{J})_\alpha, & S_{\alpha\beta} & = \frac{2}{3}  (\mathbf{r}' \times \mathbf{J})_\beta x'_\alpha, & \Xi_{\alpha\beta\gamma} & = \frac{3}{4} (\mathbf{r}' \times \mathbf{J})_\gamma x'_\alpha x'_\beta,
\end{align}
We may express the distribution of current density of the bulk medium in terms of the multipole moments~\cite{Simovski2018}
\begin{equation}
    J_\alpha = -i\omega P_\alpha + \epsilon_{\alpha\beta\gamma} \nabla_\beta M_\gamma + \frac{i\omega}{2} \nabla_\beta Q_{\alpha\beta} - \frac{1}{2} \epsilon_{\alpha\beta\delta} \nabla_\gamma \nabla_\beta S_{\delta \gamma} - \frac{i\omega}{6}\nabla_\gamma \nabla_\beta O_{\alpha\beta\gamma} +\frac{1}{6} \epsilon_{\alpha\beta\delta} \nabla_\gamma \nabla_\eta \nabla_\beta \Xi_{\delta \gamma\eta}\,
\label{currentmultipolexp}\end{equation}
where $\epsilon^{\alpha\beta\gamma}$ is the Levi-Civita symbol. Since the constitutive relations of the medium may be derived in the $\omega$-domain
\begin{align}
    \nabla \times \mathbf{E} & = i \omega \mathbf{B}, & \nabla \times \mathbf{B} & = \mu_0 (\mathbf{J}- i\omega \varepsilon_0 \mathbf{E}),
\end{align}
we can write 
\begin{equation}
    \nabla \times (\mu_0^{-1} \mathbf{B}-\mathbf{M}+\frac{1}{2}\nabla \cdot \overline{\overline{S}} -\frac{1}{6}\nabla\cdot\nabla\cdot\overline{\overline{\overline{\Xi}}})=- i\omega (\varepsilon_0 \mathbf{E}+\mathbf{P}- \frac{1}{2}\nabla\cdot \overline{\overline{Q}} + \frac{1}{6} \nabla\cdot\nabla\cdot \overline{\overline{\overline{O}}}).
\end{equation}
From which we identify the constitutive relations

\begin{eqnarray}
    \mathbf{D} & = & \varepsilon_0 \mathbf{E}+\mathbf{P}- \frac{1}{2}\nabla\cdot \overline{\overline{Q}} + \frac{1}{6} \nabla\cdot\nabla\cdot \overline{\overline{\overline{O}}}, \label{auxiliaryDdispersive}\\
    \mathbf{H} & = & \mu_0^{-1} \mathbf{B}-\mathbf{M}+\frac{1}{2}\nabla \cdot \overline{\overline{S}} -\frac{1}{6}\nabla\cdot\nabla\cdot\overline{\overline{\overline{\Xi}}}. \label{auxiliaryHdispersive} 
\end{eqnarray}
The expression in CGS units is written in the main text by applying suitable conversion formulas.

\section{Supplemental Note 8: Spatially dispersive media with octupolar responses}

In this section we introduce weak spatial dispersion to the multipolar expansion up to octupolar moments with components related to the electric and magnetic fields as in Ref.~\cite{Simovski2018}. Spatial dispersion of electromagnetic response means that the polarization current density at the observation point $\mathbf{r}$ is determined not only by the field in a given point, but also by the fields in the other areas of space.

 Considering a linear response we can write the general form for the current density $\mathbf{J}$ as the convolution of the electric field $\mathbf{E}$ and magnetic field $\mathbf{B}$ with the polarization current responses of the material $\mathcal{K}_{\alpha\beta}$

\begin{equation}
    J_\alpha(\mathbf{r}) = \int_\Omega dV' [\mathcal{K}^{\text{elec}}_{\alpha\beta}(\mathbf{r}-\mathbf{r}')E_\beta(\mathbf{r}') + \mathcal{K}^{\text{mag}}_{\alpha\beta}(\mathbf{r}-\mathbf{r}')B_\beta(\mathbf{r}') ],
\label{spatialdispersionJ}\end{equation}
where $\mathcal{K}^{\text{elec}}_{\alpha\beta}$ and $\mathcal{K}^{\text{mag}}_{\alpha\beta}$ are the components of the electric and magnetic polarization current response, and the integration is over the volume $\Omega$ around the observation point $\mathbf{r}$ that is equal to the size of the particle that constitutes the metamaterial. It is evident that Eq. (\ref{spatialdispersionJ}) can be written just in terms of spatially varying electric fields since $\mathbf{B}= 1/(i\omega) \nabla \times \mathbf{E}$. However, we will stick with this generalized definition in order to distinguish the magnetoelectric polarizations. The fields at any point inside the $\Omega$ can be expanded into a Taylor series around the center
\begin{eqnarray}
    E_\alpha (\mathbf{r}') & = & E_\alpha (\mathbf{r}) + (\mathbf{r}-\mathbf{r}')_\beta \nabla_\beta E_\alpha(\mathbf{r}) + \frac{1}{2} (\mathbf{r}-\mathbf{r}')_\beta (\mathbf{r}-\mathbf{r}')_\gamma \nabla_\beta \nabla_\gamma B_\alpha (\mathbf{r}) + \cdots \label{taylorelectricexp1}\\
    B_\alpha (\mathbf{r}') & = & B_\alpha (\mathbf{r}) + (\mathbf{r}-\mathbf{r}')_\beta \nabla_\beta B_\alpha(\mathbf{r}) + \frac{1}{2} (\mathbf{r}-\mathbf{r}')_\beta (\mathbf{r}-\mathbf{r}')_\gamma \nabla_\beta \nabla_\gamma B_\alpha (\mathbf{r}) + \cdots
\label{taylorelectricexp2}    
\end{eqnarray}
Upon inserting Eq. (\ref{taylorelectricexp1}) and (\ref{taylorelectricexp2}) into Eq. (\ref{spatialdispersionJ}), we deduce

\begin{equation}
\begin{split}
    J_\alpha(\mathbf{r}) & = a_{\alpha\beta} E_\beta(\mathbf{r}) +a_{\alpha\beta\gamma} \nabla_\gamma E_\beta(\mathbf{r}) + a_{\alpha\beta\gamma\sigma}\nabla_\gamma\nabla_\sigma E_\beta(\mathbf{r}) + \cdots \\
    & \; + b_{\alpha\beta} B_\beta(\mathbf{r}) +b_{\alpha\beta\gamma} \nabla_\gamma B_\beta(\mathbf{r}) + b_{\alpha\beta\gamma \sigma}\nabla_\gamma\nabla_\sigma B_\beta(\mathbf{r}) + \cdots 
\label{spatialdispcurrent}
\end{split}
\end{equation}
where in terms of $\mathbf{R} = (\mathbf{r}-\mathbf{r}')$, the coefficients are
\begin{align}
    a_{\alpha\beta} & = \int \mathcal\mathcal{K}^{\text{elec}}_{\alpha\beta}(\mathbf{R})dV', &  a_{\alpha\beta\gamma} & = \int \mathcal\mathcal{K}^{\text{elec}}_{\alpha\beta}(\mathbf{R})\mathbf{R}_\gamma dV', & a_{\alpha\beta\gamma \sigma} & = \int \mathcal\mathcal{K}^{\text{elec}}_{\alpha\beta}(\mathbf{R})\mathbf{R}_\gamma\mathbf{R}_\sigma dV', \\
    b_{\alpha\beta} & = \int \mathcal\mathcal{K}^{\text{mag}}_{\alpha\beta}(\mathbf{R})dV', &  b_{\alpha\beta\gamma} & = \int \mathcal\mathcal{K}^{\text{mag}}_{\alpha\beta}(\mathbf{R})\mathbf{R}_\gamma dV', & b_{\alpha\beta\gamma \sigma} & = \int \mathcal\mathcal{K}^{\text{mag}}_{\alpha\beta}(\mathbf{R})\mathbf{R}_\gamma\mathbf{R}_\beta dV'.
\end{align}
 Now, in order to compare Eq. (\ref{spatialdispcurrent}) with Eq. (\ref{currentmultipolexp}), we write the respective expansions for each of the multipole magnoelectric polarizabilities
\begin{eqnarray}
     P_i & = & p_{\alpha\beta} E_\beta+\frac{1}{2}p_{\alpha\beta\gamma} \nabla_\gamma E_\beta + \frac{1}{6}p_{\alpha\beta\gamma \sigma}\nabla_\gamma\nabla_\sigma E_\beta + \cdots + p^m_{\alpha\beta} {B}_\beta + \frac{1}{2}p^m_{\alpha\beta\gamma}\nabla_\gamma {B}_\beta +\frac{1}{6}p^m_{\alpha\beta\gamma \sigma}\nabla_\gamma\nabla_\sigma {B}_\beta \label{dipolemomentexpansion}\\
     M_i & = & m_{\alpha\beta} B_\beta+\frac{1}{2}m_{\alpha\beta\gamma} \nabla_\gamma B_\beta + \frac{1}{6}m_{\alpha\beta\gamma \sigma}\nabla_\gamma\nabla_\sigma B_\beta +\cdots  m^e_{\alpha\beta} {E}_\beta + \frac{1}{2} m^e_{\alpha\beta\gamma}\nabla_\gamma {E}_\beta + \frac{1}{6} m^e_{\alpha\beta\gamma \sigma}\nabla_\gamma\nabla_\sigma {E}_\beta\\
     Q_{\alpha\beta} & = & q_{\alpha\beta\gamma} E_\gamma+\frac{1}{6}q_{\alpha\beta\gamma \sigma} \nabla_\sigma E_\gamma +\cdots +  q^m_{\alpha\beta\gamma} {B}_l + \frac{1}{2} q^m_{\alpha\beta\gamma \sigma}\nabla_\sigma {B}_\gamma +\cdots \\
     S_{\alpha\beta} & = & s_{\alpha\beta\gamma} B_l+\frac{1}{2}s_{\alpha\beta\gamma \sigma} \nabla_\sigma B_\gamma +\cdots + s^e_{\alpha\beta\gamma}{E}_\gamma + \frac{1}{2} s^e_{\alpha\beta\gamma \sigma}\nabla_\sigma {E}_\gamma +\cdots  \\
     O_{\alpha\beta\gamma} & = & o_{\alpha\beta\gamma \sigma}E_\sigma +\cdots +  \vphantom{\frac{1}{2}}o^m_{\alpha\beta\gamma \sigma}{B}_\sigma +\cdots \\
     \Xi_{\alpha\beta\gamma} & = & \xi_{\alpha\beta\gamma \sigma} B_\sigma +\cdots + \vphantom{\frac{1}{2}}\xi^e_{\alpha\beta\gamma \sigma} {E}_\sigma +\cdots\label{octupolemagmomentexpansion}
\end{eqnarray}
where we have not written the dependence of the multipoles on $\mathbf{r}$, explicitly. Then, we substitute Eqs. (\ref{dipolemomentexpansion})-(\ref{octupolemagmomentexpansion}) into Eqs. (\ref{auxiliaryDdispersive}) and (\ref{auxiliaryHdispersive}), which yields for $\mathbf{D}$

\begin{equation}
        D_\alpha = \varepsilon_{\alpha\beta} E_\beta + \Omega_{\alpha\beta\gamma}\nabla_\beta E_\gamma + \Lambda_{\alpha\beta\gamma\sigma}\nabla_\beta\nabla_\gamma E_\sigma +p^m_{\alpha\beta} B_\beta + \Omega^m_{\alpha\beta\gamma}\nabla_\beta B_\gamma + \Lambda^m_{\alpha\beta\gamma\sigma}\nabla_\beta\nabla_\gamma B_\sigma
        \label{auxiliaryDexpanded}
\end{equation}
where $\varepsilon_{\alpha\beta} = \varepsilon_0\delta_{\alpha\beta}+ p_{\alpha\beta}$, and 

\begin{align}
\Omega_{\alpha\beta\gamma} & = \frac{1}{2}\left(p_{\alpha\gamma\beta} - q_{\alpha\beta\gamma}\right), & \Lambda_{\alpha\beta\gamma\sigma} & =  \frac{1}{12}\left( 2p_{\alpha\sigma\beta\gamma} -3q_{\alpha\beta\sigma\gamma} + 2o_{\alpha\beta\gamma\sigma}\right) \\
\Omega^m_{\alpha\beta\gamma} & = \frac{1}{2}\left(p^m_{\alpha\gamma\beta} - q^m_{\alpha\beta\gamma}\right), & \Lambda^m_{\alpha\beta\gamma\sigma} & =  \frac{1}{12}\left( 2p^m_{\alpha\sigma\beta\gamma} -3q^m_{\alpha\beta\sigma\gamma} + 2o^m_{\alpha\beta\gamma\sigma}\right) \label{lambdatermmagnetic}
\end{align}
For the auxiliary field $\mathbf{H}$, we have
\begin{equation}
    H_\alpha = \mu^{-1}_{\alpha\beta} B_\beta  + \Theta_{\alpha\beta\gamma}\nabla_\beta B_\gamma + \Pi_{\alpha\beta\gamma\sigma} \nabla_\beta\nabla_\gamma B_\sigma -m^e_{\alpha\beta}E_\beta+  \Theta^e_{\alpha\beta\gamma}\nabla_\beta E_\gamma + \Pi^e_{\alpha\beta\gamma\sigma} \nabla_\beta\nabla_\gamma E_\sigma,\label{auxiliaryHexpanded}
\end{equation}
where $\mu^{-1}_{\alpha\beta} = \mu^{-1}_0\delta_{\alpha\beta}- m_{\alpha\beta}$, and 

\begin{align}
\Theta_{\alpha\beta\gamma} & = \frac{1}{2}\left(s_{\alpha\beta\gamma} - m_{\alpha\gamma\beta}\right), & \Pi_{\alpha\beta\gamma\sigma} & = - \frac{1}{12}\left( 2m_{\alpha\sigma\beta\gamma} -3s_{\alpha\beta\sigma\gamma} + 2\xi_{\alpha\beta\gamma\sigma}\right) \\
\Theta^e_{\alpha\beta\gamma} & = \frac{1}{2}\left(s^e_{\alpha\beta\gamma} - m^e_{\alpha\gamma\beta}\right), & \Pi^e_{\alpha\beta\gamma\sigma} & = - \frac{1}{12}\left( 2m^e_{\alpha\sigma\beta\gamma} -3s^e_{\alpha\beta\sigma\gamma} + 2\xi^e_{\alpha\beta\gamma\sigma}\right)\label{Pitermelectric}
\end{align}
We are specially interested in the coefficients of the last terms in Eqs. (\ref{auxiliaryDexpanded}) and (\ref{auxiliaryHexpanded}), $\Lambda^m$ and $\Pi^e$, because they are responsible for obtaining the desired $\psi$ response,  

\begin{eqnarray}
  \mathbf{D} & = & \varepsilon \mathbf{E} + \psi \nabla^2 \mathbf{B},\\
  \mathbf{H} & = & \mu^{-1} \mathbf{B} - \psi \nabla^2 \mathbf{E},
\end{eqnarray}
in terms of magnetic dipoles, quadrupoles and octupoles moments. Therefore, in order to engineer such response it is necessary to enhance the contribution of the multipole moments in the terms $\Lambda^m$ and $\Pi^m$ from Eq. (\ref{lambdatermmagnetic}) and (\ref{Pitermelectric}).

\section{Supplemental Note 9: Quantum mechanical symmetries of polarizability tensors }

Generally, all relevant polarizabilities of a given meta-atom can be computed numerically given the distribution of polarization extracted from numerical simulations. However, to get a deeper insight into the underlying physics, it is also possible to exploit the information on the eigenmodes of the meta-atom (so-called quasi-normal modes) and construct the relevant polarizabilities using the frequencies and the structure of those eigenmodes. This calculation is fully classical, but quite technical due to the number of special features of quasi-normal modes such as their exponential divergence at large distances.

To avoid those difficulties, but still obtain the qualitiative understanding, we ignore the effect of losses. In such case, we can utilize the expressions derived in the field of molecular light scattering conveniently written in quantum-mechanical notations and summarized below.

Below, we derive some symmetry properties of the multipole moments and associated polarizability tensors. It is known that radiation fields can induce oscillating electric and magnetic multipole moments in a molecule. These moments are characterized by the electric and magnetic field components of the radiation field in the molecular property tensors~\cite{Barron2004}. By taking the expectation values of the multipole moment operators, Eqs. (\ref{dipolemomentexpansion})-(\ref{octupolemagmomentexpansion}), with the molecular perturbed wavefunctions by the radiation field in the Barron-Gray gauge, one can write the oscillating induced electric and magnetic multipole moments of the molecule in the $n$th eigenstate~\cite{Figueiredo1981}. We consider the dynamic molecular polarizability tensors to be real. The electric polarizabilities are given by \\

\begin{eqnarray}
    p_{\alpha\beta} & = & \frac{2}{\hbar} \sum_j Z_{j n} \omega_{j n} \text{Re} \braket{n|P_\alpha|j}\braket{j|P_{\beta}|n} = p_{\beta \alpha} \\
    p_{\alpha\beta\gamma} & = & \frac{2}{\hbar}\sum_j Z_{j n} \omega_{j n} \text{Re} \braket{n|P_\alpha|j}\braket{j|Q_{\beta\gamma}|n} = p_{\alpha\gamma\beta} \\
    q_{\alpha\beta\gamma} & = & \frac{2}{\hbar} \sum_j Z_{j n} \omega_{j n} \text{Re}\braket{n|Q_{\alpha\beta}|j}\braket{j|P_{\gamma}|n} =p_{\gamma\alpha\beta} \\
    p_{\alpha\beta\gamma \sigma} & = & \frac{2}{\hbar}\sum_j Z_{j n} \omega_{j n} \text{Re} \braket{n|P_\alpha|j}\braket{j|O_{\beta\gamma\sigma}|n} \\
    o_{\alpha\beta\gamma \sigma} & = & \frac{2}{\hbar}\sum_j Z_{j n} \omega_{j n} \text{Re} \braket{n|O_{\alpha\beta\gamma}|j}\braket{j|P_{\sigma}|n} = p_{\sigma\alpha\beta\gamma} \\
     q_{\alpha\beta\gamma \sigma} & = & \frac{2}{\hbar}\sum_j Z_{j n} \omega_{j n} \text{Re} \braket{n|Q_{\alpha\beta}|j}\braket{j|Q_{\gamma\sigma}|n} = q_{\gamma\sigma\alpha\beta} 
\end{eqnarray}
and for the product of magnetic and electric multipole transition moments

\begin{eqnarray}
p^m_{\alpha\beta} & = & \frac{2}{\hbar} \sum_j Z_{j n} \omega_{j n} \text{Re}\braket{n|P_{\alpha}|j}\braket{j|M_{\beta}|n}\\
m^e_{\alpha\beta} & = & \frac{2}{\hbar} \sum_j Z_{j n} \omega_{j n} \text{Re}\braket{n|M_{\alpha}|j}\braket{j|P_{\beta}|n} = p^m_{\beta\alpha} \\
      p^m_{\alpha\beta\gamma} & = & \frac{2}{\hbar} \sum_j Z_{j n} \omega_{j n} \text{Re}\braket{n|P_{\alpha}|j}\braket{j|S_{\beta\gamma}|n} \\
      s^e_{\alpha\beta\gamma} & = & \frac{2}{\hbar} \sum_j Z_{j n} \omega_{j n} \text{Re}\braket{n|S_{\alpha\beta}|j}\braket{j|P_{\gamma}|n} = p^m_{\gamma\alpha\beta}\\
      q^m_{\alpha\beta\gamma} & = & \frac{2}{\hbar} \sum_j Z_{j n} \omega_{j n} \text{Re}\braket{n|Q_{\alpha\beta}|j}\braket{j|M_{\gamma}|n} \\
      m^e_{\alpha\beta\gamma} & = & \frac{2}{\hbar} \sum_j Z_{j n} \omega_{j n} \text{Re}\braket{n|M_{\alpha}|j}\braket{j|Q_{\beta\gamma}|n} = q^m_{\gamma\alpha\beta} \\
      q^m_{\alpha\beta\gamma \sigma} & = & \frac{2}{\hbar} \sum_j Z_{j n} \omega_{j n} \text{Re}\braket{n|Q_{\alpha\beta}|j}\braket{j|S_{lm}|n}  \\
      s^e_{\alpha\beta\gamma \sigma} & = & \frac{2}{\hbar} \sum_j Z_{j n} \omega_{j n} \text{Re}\braket{n|S_{\alpha\beta}|j}\braket{j|Q_{\gamma\sigma}|n} = q^m_{\gamma \sigma \alpha\beta} \label{multipolemomentqq}\\
      p^m_{\alpha\beta\gamma \sigma} & = & \frac{2}{\hbar} \sum_j Z_{j n} \omega_{j n} \text{Re}\braket{n|P_{\alpha}|j}\braket{j|\Xi_{\beta\gamma\sigma}|n} \\
      \xi^e_{\alpha\beta\gamma \sigma} & = & \frac{2}{\hbar} \sum_j Z_{j n} \omega_{j n} \text{Re}\braket{n|\Xi_{\alpha\beta\gamma}|j}\braket{j|P_{\sigma}|n} = p^m_{\sigma \alpha\beta\gamma}\label{multipolemomentedoct}\\
      m^e_{\alpha\beta\gamma \sigma} & = & \frac{2}{\hbar} \sum_j Z_{j n} \omega_{j n} \text{Re}\braket{n|M_{\alpha}|j}\braket{j|O_{\beta\gamma\sigma}|n} \\
      o^m_{\alpha\beta\gamma \sigma} & = & \frac{2}{\hbar} \sum_j Z_{j n} \omega_{j n} \text{Re}\braket{n|O_{\alpha\beta\gamma}|j}\braket{j|M_{\sigma}|n} = m^e_{\sigma \alpha\beta\gamma }\label{multipolemomentmdoct}
\end{eqnarray}
where $\omega_{j n} = (E_j-E_n)/\hbar$ is the eigenfrequency difference between the $n$th and the $j$th state, $Z_{j n}= (\omega^2_{j n}-\omega^2)^{-1}$, and $\omega$ is the frequency of the plane-wave radiating field. In this expression we have only considered that the polarizability tensors are real, henceforth ignoring effects like absorption that can appear around the resonant frequencies $\omega \approx \omega_{jn}$. It is evident that the interference between the electric $p^m_{\alpha\beta}$ and magnetic $m^e_{\alpha\beta}$ transition dipoles is responsible for the magnetoelectric resonance which corresponds to the Tellegen medium. From the symmetry properties of the polarizability tensors we can see that in order to obtain the $\psi$-type response we need to induce oscillating multipole moments in Eqs. (\ref{multipolemomentqq}),(\ref{multipolemomentedoct}), and (\ref{multipolemomentmdoct}). To achieve this we argue that we have to overlap the resonances of magnetic dipole-electric octupole $o^m_{\alpha\sigma\beta\gamma} = m^e_{\sigma\alpha\beta\gamma}$, electric dipole-magnetic octupole
$\xi^e_{\alpha\beta\gamma\sigma}=p^m_{\sigma\alpha\beta\gamma}$, or electric and magnetic-quadrupole moments
$s^e_{\alpha\beta\gamma\sigma} = q^m_{\gamma\sigma\alpha\beta}$.

\begin{eqnarray}
      s^e_{\alpha\beta\gamma \sigma} & = & \frac{2}{\hbar} \sum_j Z_{j n} \omega_{j n} \text{Re}\braket{n|S_{\alpha\beta}|j}\braket{j|Q_{\gamma\sigma}|n} = q^m_{\gamma \sigma \alpha\beta} \\
      \xi^e_{\alpha\beta\gamma \sigma} & = & \frac{2}{\hbar} \sum_j Z_{j n} \omega_{j n} \text{Re}\braket{n|\Xi_{\alpha\beta\gamma}|j}\braket{j|P_{\sigma}|n} = p^m_{\sigma \alpha\beta\gamma}\\
      o^m_{\alpha\beta\gamma \sigma} & = & \frac{2}{\hbar} \sum_j Z_{j n} \omega_{j n} \text{Re}\braket{n|O_{\alpha\beta\gamma}|j}\braket{j|M_{\sigma}|n} = m^e_{\sigma \alpha\beta\gamma }
\end{eqnarray}

One of these options is explored and demonstrated in the manuscript main text, other options are also feasible.